\documentclass{aa}  

\usepackage{graphicx}
\usepackage{txfonts}
\usepackage[hidelinks]{hyperref}
\hypersetup{
colorlinks=true,
linkcolor=blue,
filecolor=blue,
citecolor = blue,
urlcolor=cyan,
}
\usepackage{booktabs}
\usepackage{amsmath}
\usepackage{subcaption}
\usepackage{soul}
\usepackage[commentmarkup=uwave]{changes}
\definechangesauthor[name=Gio, color=red]{gio}
\definechangesauthor[name=Ste, color=violet]{ste}
\definechangesauthor[name=Ste, color=magenta]{nb}
\definechangesauthor[name=Giada, color=teal]{giada}
\usepackage{xcolor}

\newcommand\ste[1]{{\color[rgb]{0, 0.0, 0} #1}}

\begin{document} 

   \title{Contribution of young massive stellar clusters to the Galactic diffuse $\gamma$-ray emission.}
   \authorrunning{Menchiari et al.}
   \titlerunning{Contribution of YMSCs to Galactic diffuse $\gamma$-ray emission}


   \author{S. Menchiari
          \inst{1}\fnmsep\inst{2}\fnmsep\thanks{\email{stefano.menchiari@inaf.it}}, G. Morlino\inst{1}, E. Amato\inst{1}, N. Bucciantini\inst{1}, G. Peron\inst{1} \and G. Sacco\inst{1}
          }

   \institute{$^{1}$INAF - Osservatorio Astrofisico di Arcetri, Largo Enrico Fermi 5, Firenze, Italy \\
   $^{2}$Università degli Studi di Siena, Via Roma 56, Siena, Italy }

   \date{Received 06/05/2024; accepted 20/01/25}

 
  \abstract
   {Young massive stellar clusters (YMSCs) have emerged as potential $\gamma$-ray sources after the recent association of a dozen YMSCs with extended $\gamma$-ray emission. The large size of the detected halos, comparable to that of the wind-blown bubble expected around YMSCs, makes the $\gamma$-ray detection of individual YMSCs rather challenging. As a result, the emission from most of the Galactic YMSCs could be unresolved, thus contributing to the diffuse $\gamma$-ray radiation observed along the Galactic Plane.}
   {In this study, we estimate the possible contribution to the Galactic diffuse $\gamma$-ray emission from a synthetic population of YMSCs, and we compare it with  observations obtained with different experiments, from 1 GeV to hundreds of TeV, in three regions of the Galactic Plane.}
   {As the population of galactic YMSCs is only known locally, we evaluate the contribution of $\gamma$-ray emission relying on the simulation of synthetic populations of YMSCs based on the observed properties of local clusters. We compute the $\gamma$-ray emission from each cluster assuming that the radiation is purely hadronic in nature and produced by cosmic rays accelerated at the cluster's collective wind termination shock.}
   {We find that the $\gamma$-ray emission from unresolved YMSCs can significantly contribute to the observed Galactic diffuse flux, especially in the inner part of the Galaxy, and that an important role is played by kinetic power injected by the Wolf-Rayet stellar winds.
   The predicted $\gamma$-ray flux should be considered as a lower limit, given that our calculation does not include the contribution of supernovae exploding in YMSCs.}
   {}


   \maketitle
%

\section{Introduction}
\label{sec:intro}
Stellar clusters represent the fundamental building blocks of galaxies and are among the most intensively studied celestial objects in the Cosmos. Among all clusters, the youngest and most massive ones harbor hundreds of OB-type stars, whose violent winds create the perfect environment for the production of cosmic rays (CRs). Indeed, since the '80s, it has been speculated that particle acceleration may occur in these environments, and several different scenarios have been proposed, such as acceleration by stellar wind-wind interaction \citep{Cesarsky_GammaRayStellarWinds_1983, Klepach_WindWindInteractionCR_2000, Reimer_CRfromWWinteraction_2006}, acceleration by supersonic turbulence \citep{Bykov_MSCTurbAcc_2020}, and acceleration at the collective cluster wind termination shock (TS) \citep{Morlino_2021}. Moreover, the large majority of massive stars ($\sim 96\%$) are believed to be born in clusters \citep{Parker-Goodwin:2007}, even if a non-negligible fraction ($20-30\%$) is expelled from the parent cluster due to dynamical effects \citep{Gies_ORunawayStars_1987, Parker_OstarClusters_2007}. Hence, most supernova explosions are expected to happen within a cluster. In such a scenario, particle acceleration can also occur thanks to a combination of winds and supernova explosions \citep{Vieu_MSC+SNR_2022}. 

Eventually, the detection, in the last decade, of diffuse $\gamma$-ray emission in coincidence with a dozen young massive star clusters (YMSCs), prominent among which Cygnus OB2 \citep{Bartoli_CygOB2Argo_2014, Abeysekara_CygOB2HAWC_2021, Astiasarain_CygCocoon_2023, LHAASO_UHEgammaSuperPeV_2023}, Westerlund 1 \citep{Abramowski_Wd1VHE_2012, Aharonian_HESSWesterlund1_2022} and Westerlund 2 \citep{HESS_Wd2_2011, Yang_Wd2Fermi_2018}, has further strengthened the hypothesis that these objects are efficient CR factories. In fact, the observed emissions can be easily explained if $\sim 1-10 \%$ of the power supplied by the stellar winds is used to accelerate CRs \citep{Aharonian_MSCs_2019, Peron_WindContrGalCR_2024}.

In general, the size of the detected $\gamma$-ray emission is of the order of $1^\circ \--3^\circ$, but in some cases it can be extremely large, as in the case of the 100~deg$^2$ emission observed by LHAASO around Cygnus OB2 \citep{LHAASO_UHEgammaSuperPeV_2023}. 
Noticeably, in the well established cases, the size of the emitting region is close to that expected for the wind-blown bubbles thought to surround such objects. 
The similarity can be easily explained if the accelerated particles remain mostly confined within the turbulent bubble, so that the $\gamma$-ray emission traces the projected radius of the latter.

The detection of large extended sources in $\gamma$-ray astronomy is often a challenging 
task due to the limited resolution of the telescopes, which prevents a clear disentanglement of multiple sources in crowded regions of the Galactic Plane. In addition, the background subtraction is often problematic due to the relatively small field of view of most current instruments. These observational limitations make the detection of individual YMSCs very difficult.

At present, it is reasonable to expect that most of the emission coming from YMSCs is detected as a non-resolved contribution to the large-scale Galactic emission. Alongside, recent analysis of the Galactic diffuse $\gamma$-ray emission has highlighted the presence of an excess above a few GeV up to PeV energies which is possibly associated to emission from an unresolved population of sources \citep{Zhang_LHAASOandFermiLATDiffuseGamma_2023, Cao_LHAASODiffuseGamma_2023}.

In this work, we argue that the contribution from YMSCs represents a non-negligible fraction of the diffuse Galactic emission, and could easily explain the observed $\gamma$-ray excess at very-high energies. 
Since the stellar cluster population is only known within a few kpc from the Sun \citep{Cantat-Gaudin:2020}, in order to estimate such contribution, we generate multiple synthetic populations of YMSCs based on known properties of observed local clusters. For each synthetic cluster, we simulate a mock population of stars, from which we estimate the parameters of the collective cluster wind. We compute the CR content in each YMSC using the model proposed by \cite{Morlino_2021}, and we then calculate the associated $\gamma$-ray emission due to hadronic interactions only. We include in our model three possible scenarios of plasma turbulence in the wind-blown bubble, which lead to different predictions for the $\gamma$-ray spectrum.

While \ste{we are} well aware that Supernova explosions are likely to provide a sizeable contribution to the cluster kinetic energy, comparable to, or even larger than, that of stellar winds, we do not to include CR acceleration by these sources in the current calculation. This choice deserves some comment. In the first place, aiming at estimating a lower limit to the contribution of YMSCs to the diffuse $\gamma$-ray emission, we decided not to include SNRs because these could only increase our estimate and increase its uncertainty, due to the lack of a well established description of CR production in Supernova Remnants (SNRs) expanding inside a star cluster. Our choice is justified a-posteriori by our finding that even without inclusion of SNRs, and hence in the most conservative case, YMSCs can provide a sizeable contribution to the  diffuse $\gamma$-ray emission.

The paper is structured as follows: in \S~\ref{sec:Star_pop}, we describe the method for the simulation of the mock star population in each cluster, and the recipes used to model the stellar wind physics; in \S~\ref{sec:YMSC_pop}, we illustrate the adopted approach for the simulation of the synthetic YMSC population; in \S~\ref{sec:CR_model}, we summarize the model of CR acceleration at the wind TS developed by \cite{Morlino_2021} and the procedure used to compute the $\gamma$-ray emission; in \S~\ref{sec:Comparison_with_obs}, we compare the diffuse emission from the Galactic population with available data in the literature, and we then discuss the obtained results. Finally, we present our conclusions in \S~\ref{sec:conclusions}.

\section{Simulation of a mock stellar population and estimation of the cluster wind parameters.}
\label{sec:Star_pop}
For each YMSC, the mock stellar population is created based on two fundamental parameters: the cluster age ($t_{\rm sc}$) and the number of stars ($N_\star$) at the time the cluster was formed, with $N_\star$ directly linked to the cluster mass (see Appendix~\ref{app:f_nstar}). We build our populations of stars in two steps:
first, a population of $N_\star$ stars is extracted by randomly sampling the initial stellar mass function (IMF); afterward, based on the cluster age, all the stars that are expected to have exploded as supernovae are removed from the population.

We here consider the IMF provided by \cite{Kroupa_IMF_2001}. When sampling the IMF, we fix the minimum and maximum stellar masses that can be generated to $M_{\star,\min}=0.08$~M$_\odot$ and $M_{\star,\max}=150$~M$_\odot$.  $M_{\star,\min}$ is related to the minimum theoretical mass to support nuclear burning \citep{Carroll_IntroModAstro_1996}, while $M_{\star,\max}$ is the maximum observed stellar mass in clusters \citep{Zinnecker_MstarMax150_2007}\footnote{More recent works investigating the cluster R~136 infer a maximum stellar mass as large as $\sim 300\, \rm M_{\odot}$ \citep{Crowther+2010}. However, we here adopt the smaller value of 150 M$_{\odot}$ to provide a more conservative estimate for the $\gamma$-ray emission.}.

In order to remove all the stars that have exploded as supernovae from the population of stars in a cluster of given age, we adopt the following criterion:
a star of a specific mass $M_\star$ will leave the main sequence, and subsequently explode as a supernova, at a turn-off time ($t_{\rm to}$) given by \citep{Buzzoni_TOtime_2002}:
\begin{equation}
\label{eq:tTO}
\log_{10} \left(\frac{t_{\rm to}}{1 \rm \ yr} \right) = 0.825 \log_{10}^2 \left(\frac{M_{\star}}{120 \rm \ M_\odot} \right) + 6.43 \ .
\end{equation}
We do not consider any post-main-sequence evolution (with the exception of stars in the Wolf Rayet phase, that will be investigated separately below): namely, all the stars with $t_{\rm to}<t_{\rm sc}$ are removed from the cluster. We remind that, as anticipated in Section\,\ref{sec:intro}, we neglect the energy released by the supernova explosions in the calculations of the CR power.

As an additional case of investigation, we build a mock stellar population including the presence of Wolf-Rayet (WR) stars. To this purpose, we assume that all stars with $M_\star>25$~M$_\odot$ undergo a WR phase for $\sim$0.3 Myr after $t_{\rm to}$ \citep{Rosslowe_WRsAge_2015}, with the choice of the minimum WR mass motivated by the results of stellar evolution codes, showing that the minimum mass to develop a WR is in the range 22--37~M$_\odot$ \citep{Eldridge_WRMinMass_2006}. 
Lower mass WR stars might in fact be born as a result of binary evolution. Those stars may provide a sizeable contribution to the wind power, but mostly at late times. In fact, this channel is important for stars in the 10-20 $M_\odot$ range which leave the main sequence at ages $\gtrsim 10$ Myr, hence this effect is bound to have a minor impact for the clusters considered in this work. In any case, neglect of this additional contribution can only lead to underestimate the wind kinetic luminosity, and hence make our estimate of the YMSC contribution to the diffuse $\gamma$-ray background a more conservative lower limit.

\subsection{Modeling stellar winds}
The physics of stellar winds is modeled using a purely empirical approach. The mass loss rate of a given main sequence star ($\dot{M}_\star$) is calculated using the relation provided by \cite{Nieuwenhuijzen_Mdot_1990}:
\begin{equation} \label{eq:MdotNieu}
    \dot{M_\star} = 10^{-14.02} \, \left( \frac{L_\star}{\rm L_\odot} \right)^{1.24} \, \left( \frac{M_\star}{\rm M_\odot} \right)^{0.16} \, \left( \frac{R_\star}{\rm R_\odot} \right)^{0.81 } \, \frac{\rm M_\odot}{\rm yr}\,.
\end{equation}
The kinetic luminosity of the stellar winds ($L_{\star,\rm w}$) is defined as:
\begin{equation}
\label{eq:LwStar}
    L_{\star,\rm w} = \frac{1}{2} \dot{M_\star} v_{\rm w}^2 
    = \frac{1}{2} \dot{M_\star} \left  \{ C(T_{\rm eff})^2 \left [\frac{2 G M_\star (1-L_\star/L_{\rm Edd})}{R_{\star}} \right ] \right \} \ ,
\end{equation}
where $R_{\star}$ is the stellar radius, $L_{\star}$ is the stellar bolometric luminosity, $L_{\rm Edd}$ is the Eddington luminosity and the term in braces represents the wind speed squared \citep{Kudritzki_WindsHotStars_2000}. The coefficient $C(T_{\rm eff})$ depends on the stellar effective temperature $T_{\rm eff}$ and is inferred from observations \citep{Kudritzki_WindsHotStars_2000}. The stellar effective temperature is calculated using Boltzmann's law 
\begin{equation}
  T_{\rm eff}=\left [\frac{L_\star}{4 \pi R_{\star}^2 \sigma_{\rm b}} \right ]^{1/4} \,,    
\end{equation}
where $\sigma_{\rm b}$ is the Boltzmann constant. The stellar radii and luminosities are assigned based on empirical relations. For stellar radii, we use the relation proposed by \cite{Demircan_StarsMRR_1991}: $R_{\star}=0.85 \left(\frac{M_\star}{M_\odot} \right)^{0.67}$ R$_\odot$. For stellar luminosities we use the broken power-law relation in Equation~(B.4) of \cite{Menchiari_CygOB2_2024}. 

When considering WR stars, we compute the mass loss rate according to \cite{Nugis_MdotWR_2000}:
\begin{equation} \label{eq:MdotRenzo}
    \dot{M}_{\star,\textrm{WR}} = 10^{-11.0} \,  
    \left (\frac{L_{\star, \textrm{WR}}}{L_\odot} \right)^{1.29} \, 
    \left( \frac{Y_{\textrm{WR}}}{Y_\odot} \right)^{1.73} \,
    \left( \frac{Z_{\textrm{WR}}}{Z_{\odot}} \right)^{0.47} \,
    \frac{\rm M_\odot}{\rm yr} \, ,
\end{equation}
where $Y_{\textrm{WR}}$ and $Z_{\textrm{WR}}$ are the helium fraction and metallicity of WR stars respectively (both normalized to solar values). $L_{\star, \textrm{WR}}$ is the bolometric luminosity of the WR, which can be calculated using an ad-hoc empirical mass-luminosity relation \cite[see Equation~(3) by][]{Schaerer_WRsMLR_1992}.
Note that since $L_{\star, \textrm{WR}}$ depends on the WR mass, it can drastically change in time due to severe mass losses associated to the wind. We account for this dependence by considering the time evolution of the stellar mass in the evaluation of the mass loss rate, i.e. $\dot{M}= \dot{M}\left[L_{\star,{\rm WR}}\left(M_{\star}(t)\right)\right]$.
Finally, the kinetic luminosity of the WR wind is calculated using Equation \eqref{eq:LwStar}, but considering a constant wind speed of 2000 km~s$^{-1}$. \ste{In general, the wind speed in WR stars is not constant and depends on the type of WR. For example, WC and WO stars have a terminal wind speed in the range $\sim 1000 \--3000$\,km$^{-1}$\,s$^{-1}$ \citep{Niedzielski_2002, Crowther_WRsReview_2007}, while WN stars have wind velocities spanning $\sim 1000 \-- 2000$\,km$^{-1}$\,s$^{-1}$ \citep{Crowther_WRsReview_2007}.} 

Once the wind luminosity and mass loss rate of every \textit{i}--th star of the cluster are known, the collective cluster wind mass loss rate ($\dot{M}$) and luminosity ($L_{\rm w}$) are obtained using mass and momentum conservation.

\section{Generating synthetic population of Galactic YMSCs}
\label{sec:YMSC_pop}
In order to simulate the Galactic YMSCs, we start from the cluster formation rate, 
\begin{equation} 
\label{eq:N_YMSC}
  \xi_{\rm sc}(M_{\rm sc}, t, r, \theta) \equiv \frac{dN_{\rm sc}}{dM_{\rm sc} \, dt \, r dr \, d\theta}\ ,
\end{equation}
which gives the number of clusters born at time $t$ per unit mass, time and surface. As commonly done in the literature, we here assume that the dependence of $\xi_{\rm sc}$ on space, time and mass, can be factorized, so that $\xi_{\rm sc}(M_{\rm sc}, t, r, \theta)= \psi(t) \, f(M_{\rm sc}) \, \rho(r, \theta)$, where $\psi(t)$, $f(M_{\rm sc})$ and $\rho(r, \theta)$ are, in order, the total cluster formation rate, the cluster initial mass function (CIMF) and the cluster spatial distribution.

\subsection{Mass distribution of star clusters}
The CIMF can be inferred from the from observation of the local SC population. In this regard, we rely on the work by \cite{Piskunov_GalSCGlobSurvIV_2018}, based on the Milky Way Star Cluster Survey, where the CIMF is modeled as a broken power law: for clusters with $M_{\rm sc} > 10^3$~M$_\odot$, the CIMF is $f(M_{\rm sc}) = 1.2 (M_{\rm sc}/M_{\odot})^{-1.54}\, \rm M_{\odot}^{-1}$. We neglect SCs with mass smaller than $10^3 \, \rm M_{\odot}$ because the integrated wind kinetic luminosity of those clusters is $< 1\%$ of the entire YMSC population. The maximum cluster mass is derived from observations \citep{Piskunov_GalSCGlobSurvIV_2018}, and is $M_{\rm sc, \max}=6.3 \times 10^4$ M$_\odot$, corresponding to the mass of Westerlund~1. Notice that in generating the mock SC distribution, the number of stars is used in place of the cluster mass, see Appendix~\ref{app:f_nstar}.

\subsection{Age distribution of star clusters}
\begin{figure}
\centering
     \begin{subfigure}[b]{\columnwidth}
         \centering
         \includegraphics[width=\columnwidth]{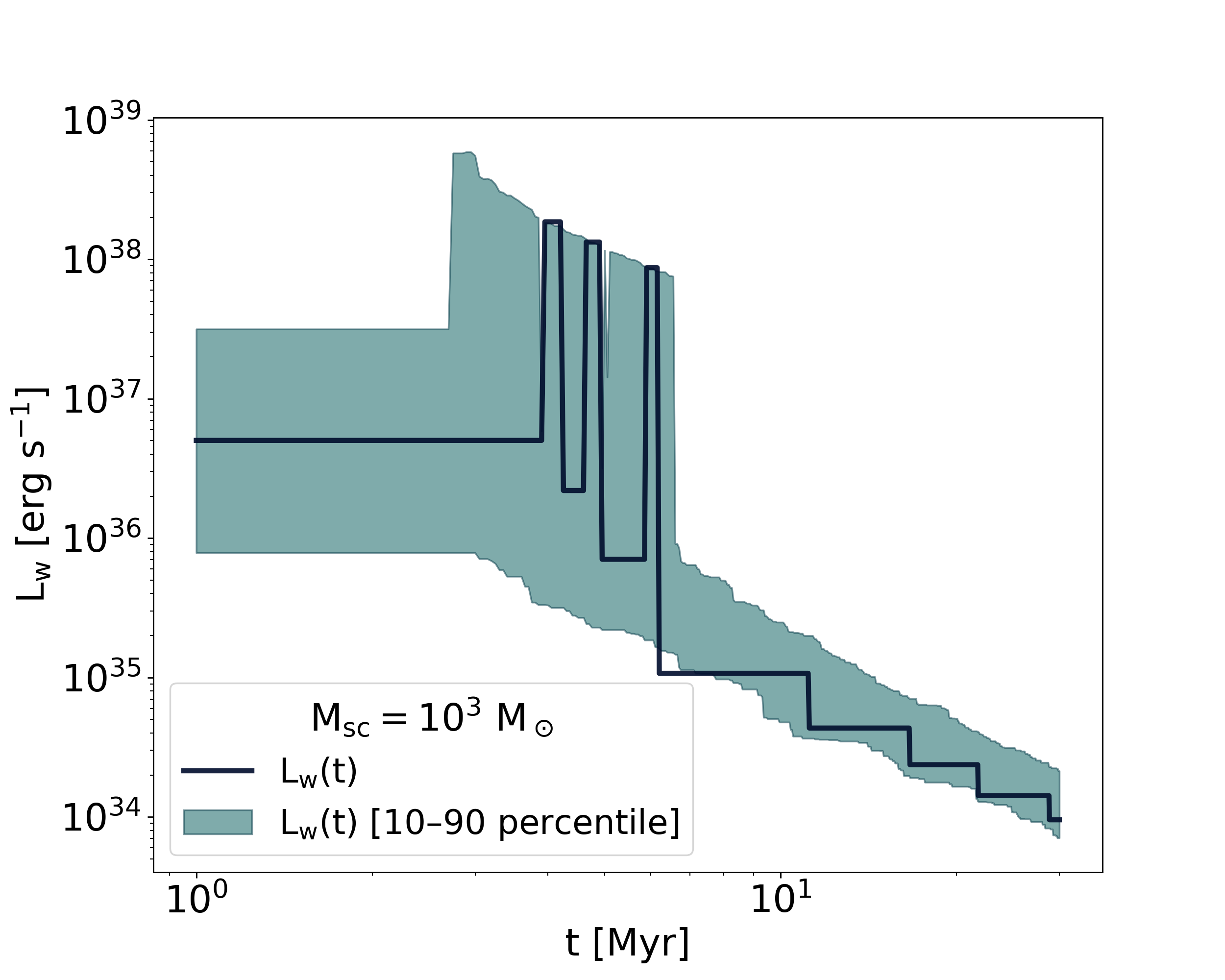}
     \end{subfigure}
     \begin{subfigure}[b]{\columnwidth}
         \centering
         \includegraphics[width=\columnwidth]{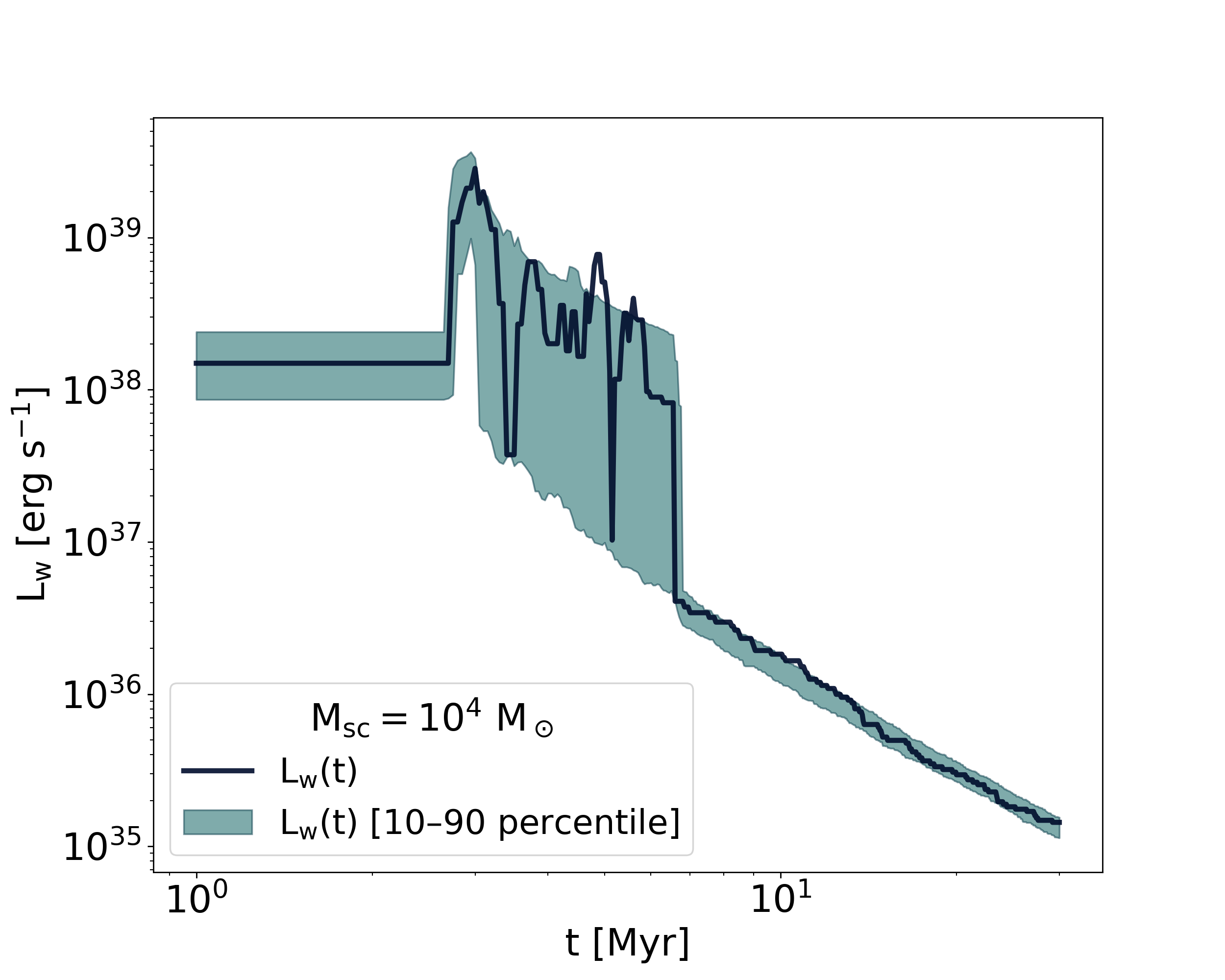}
     \end{subfigure}
\caption{The wind power as a function of time for a YMSC with 10$^3$\,M$_\odot$ (upper plot) and 10$^4$\,M$_\odot$ (lower plot) is shown as a solid line. The shaded region shows how the wind luminosity can statistically vary for different random sampling of the initial stellar mass function. The limits of the shaded region are calculated as the 10\% and 90\% percentile out of 100 possible samplings of the initial stellar mass function.}
\label{fig:Lw_vs_t}
\end{figure}

Similarly to the CIMF, the SCs age distribution can be retrieved from the local cluster population. As we are interested in YMSCs, we consider only star clusters with age between $t_{\rm sc, \min}=0$ and $t_{\rm sc, \max}=10$ Myr, because the wind luminosity drops substantially at later times. This can be seen in Figure~\ref{fig:Lw_vs_t}, where we show the wind power as a function of time for two YMSCs with masses of 10$^3$ and 10$^4$ M$_\odot$. Since the wind power depends on the number of massive stars, we also show the range in which it can vary, calculated as the 10\% and 90\% percentile out of 100 possible samplings of the stellar IMF.
After accounting for the effect of cluster evolution, the age distribution of stellar clusters is found to be approximately uniform in the last few tens of Myrs \citep{Piskunov_GalSCGlobSurvIV_2018}. This implies a constant cluster formation rate in the time interval we are interested in. This rate can be estimated from the star formation rate of local young stellar clusters within 1 kpc as measured by \cite{Bonatto_SFRinSC_2011}. Assuming the CIMF from \cite{Piskunov_GalSCGlobSurvIV_2018}(underfilling case) leads to an average local cluster formation rate of $\psi= 1.8$ Myr$^{-1}$ kpc$^{-2}$ \citep{Menchiari_ProbingSCsCRsPhD_2023}. 

\subsection{Spatial distribution of star clusters}
\label{app:RadialDistrib}
The spatial distribution of young clusters is strictly linked to the value of the cluster formation rate across the Milky Way. This is well traced by the position of the giant molecular clouds (GMCs), which in turn follows the spiral pattern of the Galaxy. 
We spatially allocate the synthetic stellar clusters following a two-step procedure:
\begin{enumerate}
\item We generate stellar clusters with a galactocentric radial distribution following that of GMCs (under the assumption of an isotropic angular distribution) and assuming an exponential altitude distribution similar to that of the observed gas profile.
\item Based on its radial and angular position, we associate each synthetic cluster to a specific Galactic structure, i. e. spiral arm, galactic bar, etc. We detail this in the following.   
\end{enumerate}
For the radial distribution of GMCs and the modeling of the Milky Way spiral structure, we rely on the work by \cite{Hou_MWStructure_2014}, who also supply a complete catalog of GMCs where, for each cloud, the galactic coordinates and kinematic velocity are reported. For a large portion of the GMCs, the kinematic ambiguity is resolved. We compute the kinematic distance of every cloud with disentangled kinematic ambiguity using the code developed by \cite{Wenger_KinDistCode_2018}. When doing so, we adopt the classical rotation curve based method, using the state-of-the-art curve provided by \cite{Reid_GalRotCurve_2019}. A small fraction of GMCs in the catalog have also distances calculated using more reliable methods such as parallax. We hence adopt those estimates when available. Once the position of the GMCs is known, we can compute their radial distribution, which we express in terms of the surface mass density of molecular gas ($\Sigma_{H_2}^{GMC}$). To calculate the surface mass density, we average the total GMCs mass in 18 rings with constant width of 1 kpc and radii spanning in the interval 0--18 kpc, and centered on the Galactic center. At this point we can formally define the radial distribution of YMSCs, normalized at the Sun position, as:
\begin{equation}
\label{eq:YMSCRadDistr}
\rho(r)=\frac{\Sigma_{H_2}^{GMC}(r)}{\Sigma_{H_2}^{GMC}(r=8.5 \rm \ kpc)}
\end{equation}
Notice that the normalization of $\xi_{\rm sc}$ at the Sun position is given by the local observed cluster formation rate $\psi$.

Once the radial distribution is known, we allocate stellar clusters following the Milky Way observed morphology. To reproduce the Milky Way structure we use the four logarithmic arms model provided by \cite{Hou_MWStructure_2014}, defined as:  
\begin{equation}
\label{eq:LogSpiral}
\ln \left( \frac{r}{R_i} \right) = \left(\frac{\theta-\theta_i}{1 \rm \ rad} \right) \tan \Psi_i
\end{equation}
where $R_i$, $\theta_i$, and $\Psi_i$ are parameters inferred from the fit to the position of H\textsc{II} regions, masers and GMCs. We report their values for each arms and for the local spur in Table \ref{tab:SpiralArmPar}.
\begin{table}
\begin{center}
\begin{tabular}[c]{l c c c c c}  
\toprule \toprule
Parameters  & Arm 1 & Arm 2 & Arm 3 & Arm 4 & \vtop{\hbox{\strut Local}\hbox{\strut Spur$^\star$}} \\
\midrule
R$_i$ [kpc] & 3.27 & 4.29 & 3.58 & 3.98 & 8.16\\
$\Psi_i$ [$^\circ$] & 9.87 & 10.51 & 10.01 & 8.14 & 2.71\\
$\theta_i$ [$^\circ$] & 38.5 & 189 & 215.2 & 320.1 & 50.6\\
\bottomrule 
\end{tabular}
\caption{Parameter values used in Eq.~\eqref{eq:LogSpiral} to model the Milky Way spiral arms \citep{Hou_MWStructure_2014}. $^\star$Note that the Local Spur is defined only for $50.6^\circ<\theta<110^\circ$.}
\label{tab:SpiralArmPar}
\end{center}
\end{table}  
On top of the spiral structure, we also take into account the structure of the innermost region. Here we consider the presence of the Galactic bar and the Near 3 kpc and Far 3 kpc arms. The first is modeled as an ellipse with semi-major axis of 3.3 kpc and an aspect ratio of 4:10 \citep{Churchwell_GalBarGlimpse_2009}.
The Near/Far-3-kpc Arms are modeled using an elliptical annulus, with a semi-major axis of 4.1 kpc, and an aspect ratio of 0.54 (a semi-minor axis of 2.2 kpc) \citep{Green_NF3kpcArms_2011}.

To allocate the synthetic YMSCs in the Galaxy we use the following procedure: first, for every j-th cluster, we start by randomly generating its radial ($r_j$) and angular ($\theta_j$) coordinates. Radial distances are extracted considering the probability distribution $\rho(r)$ given by Equation~\eqref{eq:YMSCRadDistr}, while the angular coordinate is chosen assuming a uniform distribution. Afterward, we randomly choose the structure (i.e. spiral arm, local spur or Near/Far-3-kpc Arms and Galactic Bar) to associate the cluster with, among the ones that satisfy the following criteria:
\begin{itemize}
    \item Spiral Arm $i=1$ to 4, if $r_j>R_i $;
    \item  Local spur if  $7.59 <r_j/\rm kpc <9.17$ and $50.6^\circ<\theta_j<110^\circ$;
    \item Spiral Arm $i=1, 3, 4$ or Near/Far-3-kpc Arms and Galactic Bar, if $R_i < r_j<4.29$ kpc;
    \item Near/Far-3-kpc Arms and Galactic Bar, if $r_j<3.27$ kpc.
\end{itemize}
When more structures satisfy the criteria, they have the same probability to be chosen. Finally, when a cluster is associated with the Near/Far-3-kpc Arms and the Galactic Bar structures, the ultimate choice between which of the two structures to be matched is determined by a criterion of minimum distance.

Note that the angular coordinate $\theta_j$ is used only to check whether the YMSC should be associated with the Local Spur. Once the cluster is placed in a given arm, the coordinate $\theta_j$ is recalculated by inverting Equation~\eqref{eq:LogSpiral}:
\begin{equation}
\theta_j= \left(\frac{\tan \Psi_i}{1 \ \rm rad} \right )^{-1} \ln \left( \frac{r_j}{R_i} \right) + \theta_i \ .
\end{equation}
When a cluster is associated with the Galactic Bar we check whether its position is actually located within the Bar. If not, $\theta_j$ is randomly extracted until its position falls within the Bar. Finally, when a cluster is associated with the Near/Far-3-kpc Arm, we change its coordinates to the nearest point belonging to the ellipse defining these structures.

Once all clusters are associated to a structure, we proceed to perturb their positions according to a Gaussian distribution, following the fundamental concept that both Spiral Arms and the Near and Far structure possess an intrinsic thickness. The width of the spiral arm increases with galactocentric distance, so that the probability $\mathcal{P}$ of finding a source displaced from the center of the arm by $\Delta r_j$, is \citep{Faucher_IsolatedRadioPulsar_2006}:
\begin{equation}
\mathcal{P}(\Delta r_j) = \frac{1}{2 \pi (0.07 r_j)} \, \exp\left[- \frac{\Delta r_j^2}{2 (0.07 r_j)^2}\right] \,.
\end{equation}
As for the clusters belonging to the Near and Far arms, we extract from a Gaussian probability distribution the Galactic x and y coordinates with a spread of $\sigma_x = \sigma_y = 0.1667$ kpc, such that the probability at $3 \sigma$ returns a scattering compatible with the observed radial thickness of 0.5 kpc \citep{Green_NF3kpcArms_2011}. Notice that the thickness of a spiral arm in YMSCs may be slightly different from that inferred for pulsars, however the  result on the diffuse emission depends very weakly on this parameter. 

At last, we generate the vertical coordinate ($z$) following the observed gas distribution, i.e. an exponential profile with a characteristic spread of 100 pc \citep{Strong_GalpropDiffGamma_2000}:
\begin{equation}
\rho(z)=\exp \left(- \frac{z}{100 \ \rm pc} \right) \,.
\end{equation} 

\begin{figure*}[t]
\begin{center}
\includegraphics[width=0.95\textwidth]{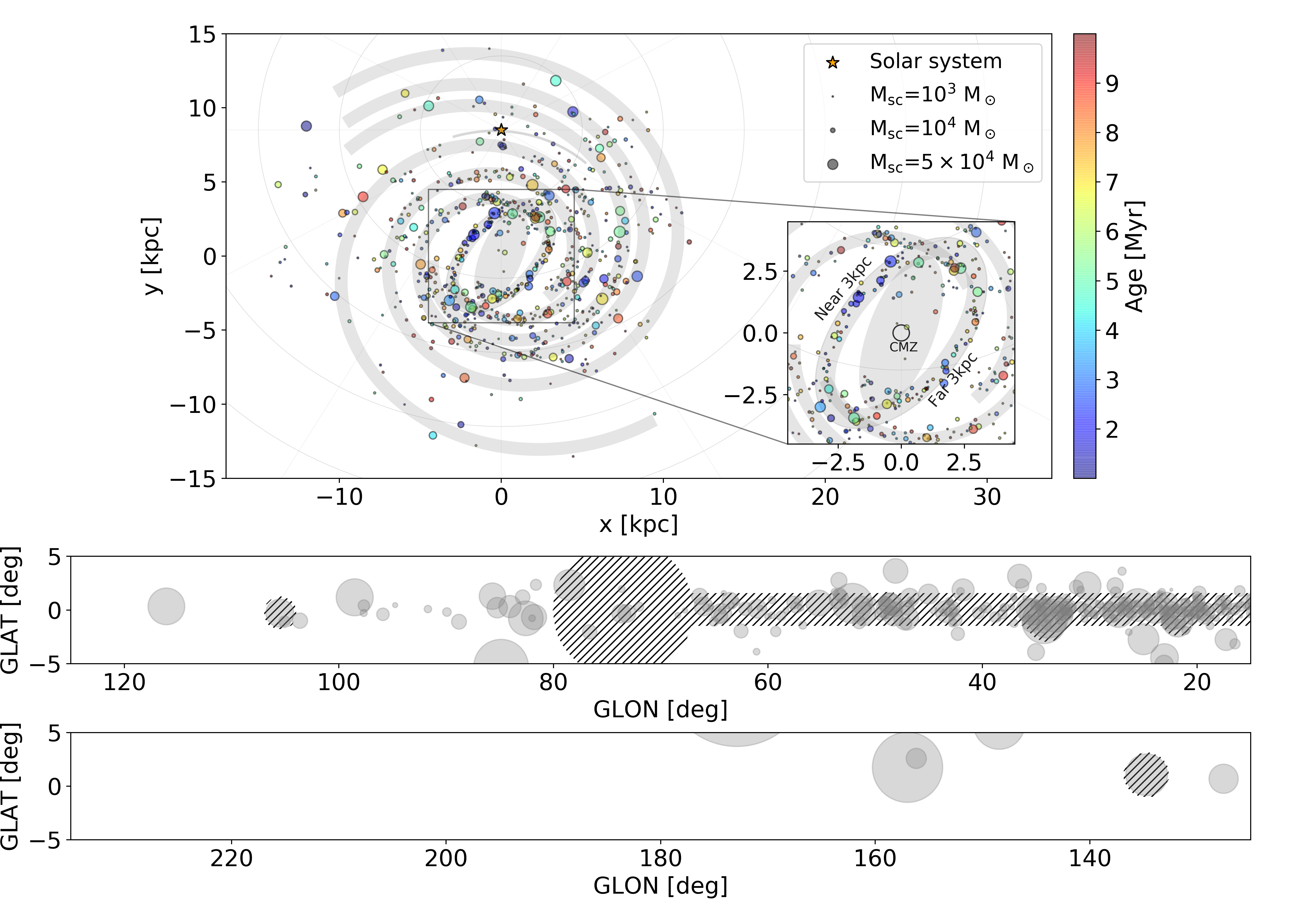} 
\caption{A single realisation of the Milky Way population of YMSCs. \textit{Top panel}: Face-on view of the Milky Way. The size and color of each YMSC refer to its mass and age respectively, according to the scales indicated in the panel. Notice that the size distribution is continuous and the associations between mass and size reported in the panel are only for reference. The grey regions represents all the Galactic structures (spiral arms, Local Spur and Galactic bar) employed in the simulation of the Milky Way cluster population. The inset shows a zoomed view of the Milky Way center. The location of the Central Molecular Zone (CMZ) is also reported for reference. The star marks the position of the Solar System. \textit{Middle and bottom panels}: Edge on view of the Milky Way towards \texttt{ROI1} (upper panel) and \texttt{ROI2} (lower panel). Grey circles represent the projected sizes of the wind-blown bubbles surrounding YMSCs. The striped regions correspond to the $\mathcal{A}\cup \mathcal{B}$ masks employed in our analysis for this specific realization of the Milky Way} (see Appendix~\ref{app:mask}).
\label{fig:SC_distrib}
\end{center}
\end{figure*}

With the cluster formation rate $\xi_{\rm sc}$ defined it is possible to finally build the synthetic SC population. We find a total number of 747 clusters with average values for the wind luminosity and the mass loss rate equal to $\langle L_{\rm w} \rangle \approx 3 \times 10^{36}$~erg~s$^{-1}$ and $\langle\dot{M}\rangle \approx 10^{-6}$~M$_\odot$ yr$^{-1}$, respectively. Figure\,\ref{fig:SC_distrib} shows a possible realization of the Milky Way YMSCs population.

\section{Cosmic ray acceleration at the collective wind termination shock of YMSCs}
\label{sec:CR_model}
When the average distance between stars in a cluster is less than the size of individual stellar wind TS, the winds from massive stars are expected to combine to create a collective cluster wind. This scenario is favored in young clusters as a direct consequence of mass segregation at birth \citep{Lada_SCReview_2003}, namely the fact that the most massive stars (hence those producing the most powerful winds) forms at the very core of the SC. In our mock SC population, the star distribution is not simulated. We simply assume that all SCs are compact enough to generate a collective TS. As the wind material is shocked and heated, it expands adiabatically in the interstellar medium (ISM) generating large bubble-like structures similar to those observed close to single massive stars \citep{Weaver_1977}. A model for particle acceleration in these systems was presented by \cite{Morlino_2021} based on the idea that particles are accelerated at the collective wind TS via the diffusive shock acceleration mechanism, and subsequently escape from the acceleration site experiencing a combination of advection and diffusion within the hot bubble, until reaching its border. From there, CRs are free to escape in the unperturbed ISM. 

Given the highly turbulent nature of the shocked wind, particle diffusion is likely suppressed, resulting in a long confinement time within the bubble where the majority of $\gamma$-ray emission is produced through hadronic interactions. The CR distribution in the bubble is \citep{Morlino_2021}:
\begin{equation}
\label{eq:FDownstream}
f(r,p)= f_{\rm ts}(p) \, \mathcal{G}(D_2, R_{\rm b}) \ ,
\end{equation}
where the distribution function at the TS is
\begin{equation}
\label{eq:fTSnorm}
 f_{\rm ts}(p) = \mathcal{K}(\epsilon_{\rm cr}, L_{\rm w}, \dot{M}) \left ( \frac{p}{m_{\rm p} c} \right )^{s} \left [ 1 + a_1 \left (\frac{p}{p_{\max}} \right )^{a_2} \right ] 
 e^{- a_3 \left(\frac{p}{p_{\max}} \right )^{a_4}} \ .
\end{equation}
The parameters $a_{1,..,4}$ are determined by the energy dependence of the diffusion coefficient, which is ultimately related to the MHD turbulence spectrum. In this work, we will consider three different diffusion regimes: Kolmogorov, Kraichnan and Bohm-like diffusion. The coefficients $a_{1,..,4}$ relevant for each of these three diffusion regimes are reported in Table~\ref{tab:fTSparamenters}. This coefficients are derived in \cite{Menchiari_CygOB2_2024} and are valid for any YMSCs. The parameter $s$ is the spectral injection slope of accelerated particles. We fix for all clusters $s=-4.2$. This value results from the modelling of the Cygnus OB2 emission performed by \cite{Menchiari_CygOB2_2024}, but a similar spectrum is also deduced from $\gamma$-ray measurements of other SCs \cite[see e.g.][]{ Aharonian_MSCs_2019,Yang2017NGC3603,Aharonian_HESSWesterlund1_2022,Liu2022M17, Mitchell+2024} for which an energy spectrum of 2.3-2.4 is found.
\begin{table}
\begin{center}
\begin{tabular}[c]{l c c c c}  
\toprule \toprule
 Models  & a$_1$ & a$_2$ & a$_3$ & a$_4$ \\
 \midrule
 Kolmogorov & 10 & 0.308653 & 22.0241 & 0.43112\\
 Kraichnan & 5 & 0.448549 & 12.52 & 0.642666\\
 Bohm & 8.94 & 1.29597 & 5.31019 & 1.13245\\
\bottomrule 
\end{tabular}
\caption{Parameters values used to calculate the distribution function in Equation~\eqref{eq:fTSnorm} for different assumption of the diffusion coefficient.}
\label{tab:fTSparamenters}
\end{center}
\end{table}
$\mathcal{K}$ is a normalization factor that depends on the cluster wind luminosity, mass loss rate and CR acceleration efficiency ($\epsilon_{\rm cr}$) \cite[see equation~(12) in][]{Menchiari_CygOB2_2024}. The function $\mathcal{G}$ \cite[see Equations~(10b) and (11) in][]{Menchiari_CygOB2_2024} depends on the diffusion coefficient in the bubble and the size of the bubble, which reads \citep{Weaver_1977}: $R_{\rm b}=0.798 \, L_{\rm w}^{1/5} \rho_0^{-1/5} t_{\rm sc}^{3/5}$, where $\rho_0$ is the average density close to the YMSC that we assume to be $10 \, m_{\rm p}$ cm$^{-3}$ ($m_{\rm p}$ is the mass of the proton). 
When calculating the diffusion coefficient in the system, we assume, for the Kolmogorov and Kraichnan cases, that the turbulence is injected at a characteristic length scale of $\sim 1$\,pc, roughly corresponding to the size of the cluster core. In the Bohm case, we assume, instead, that the injection of turbulence occurs at all scales.  
The parameter $p_{\max}$ is the maximum momentum determined based on the Hillas (confinement) criterion: $D_{\rm w}(p_{\max})/v_{\rm w}=R_{\rm ts}$, where $D_{\rm w}$ is the diffusion coefficient in the cold cluster wind, $v_{\rm w}=(2 L_{\rm w} / \dot{M})^{1/2}$ is the collective cluster wind speed and $R_{\rm ts} = 0.791 \dot{M}^{1/2} v_{\rm w}^{1/2} L_{\rm w}^{-1/5} \rho_0^{-3/10} t_{\rm sc}^{2/5}$ is the size of the TS \citep{Weaver_1977}. 

The $\gamma$-ray flux from a single YMSC due to hadronic interactions is calculated as follows:
\begin{equation}
\label{eq:PhiGammaYMSC}
\phi_{\gamma}(E_\gamma)=\frac{c n_0}{d^2} \int \int_{R_{\rm ts}}^{R_{\rm b}} r^2 f(r, E_p) \frac{d \sigma(E_p, E_\gamma)}{dE_p} dr dE_p
\end{equation}
where $d$ is the distance from Earth and $n_0=10\, \rm cm^{-3}$ is the average target density, which is expected to be close to the external ISM density, as a result of the expected fragmentation of the swept up shell \citep{Blasi-Morlino:2024}. 
The differential cross section for $\gamma$-ray production, $d\sigma/dE_p$, is taken from  \cite{Kafexhiu_SigmaPi0Gamma_2014} and based on SYBILL.

Since, the $\gamma$-ray emissivity integrated along the line of sight is expected to depend on the projected angular radius only weakly \cite[see][]{Menchiari_CygOB2_2024}, in order to speed up the calculation, we use, as a spatial model for the sources, a uniform disk of size equal to that of the projected wind bubble.

Finally, we note that the $\gamma$-ray emission depends almost linearly on the bubble volume due to the swept up mass. Hence, a correct estimate of the bubble size is important.
The adiabatic model by \cite{Weaver_1977} used here overestimates the bubble size because \ste{it} does not account for the ISM external pressure nor for the radiative losses. While the former has a very minor impact, the latter may be relevant at times $\gtrsim 10$\, Myr \citep{Silich_RadPressYMSCs_2013}. Such a timescale may be even reduced due to the external shell fragmentation which increases the contact surface between the hot shocked wind and the cold ISM \citep{Lancaster_StellarWindCooledII_2021}.
On the other hand, when also SN explosion are taken into account, the effect of cooling will be \ste{compensated} by the injection of additional power. Including all those effects is a not trivial task because it requires a full time dependent calculation able to account for the full 3D hydro-dynamical evolution of the bubble. However, at the end of Section~\ref{sec:Comparison_with_obs} we will evaluate the impact of a possible bubble size reduction by arbitrarily reducing its radius.

\section{The contribution of YMSC to diffuse $\gamma$-ray emission}
\label{sec:Comparison_with_obs}

We now focus on two regions of the sky for which recent observations of the diffuse $\gamma$-ray emission are available in the literature, and compute the corresponding emission of our synthetic YMSCs. 
The first region (named \texttt{ROI1}), is defined for  $|b|\leq 5^\circ$ and $15^\circ \leq l \leq 125^\circ$, while the second region (named \texttt{ROI2}) is defined for $|b| \leq 5^\circ$, and $125^\circ \leq l \leq 235^\circ$. For both \texttt{ROI1} and \texttt{ROI2}, flux measurements are provided by Fermi-LAT \citep{Zhang_LHAASOandFermiLATDiffuseGamma_2023} and LHAASO \citep{Cao_LHAASODiffuseGamma_2023}. In \texttt{ROI1} additional measurements of the diffuse flux are provided by the ARGO experiment\footnote{The ARGO diffuse flux is  extracted from a region slightly smaller than \texttt{ROI1} and defined as $|b|\leq 5^\circ$ and $25^\circ \leq l \leq 100^\circ$. However, given the small differences in extension and position in the sky, we decided to include ARGO data in our work.} \citep{Bartoli_ArgoDiffuse_2015}.

\subsection{Masking the contribution of resolved sources}
Notice that all the data we use are obtained after removing known sources, to avoid an artificial increase of the Galactic diffuse emission. However, the procedure used for this purpose are different for each of the considered datasets. LHAASO measurements are obtained after masking sources from both the TeVCat \citep{Wakely_TeVCat_2008} and the 1LHAASO catalog \citep{Cao_LHAASOCat_2023}, while in the ARGO analysis only TeVCat sources have been masked\footnote{In this respect, the ARGO flux can be potentially polluted by the emission from sources recently detected by LHAASO and not included in the TeVCat.}. The Fermi-LAT data results from applying the same masks used in the LHAASO analysis after removing the emission of the 4FGL sources \citep{Abdollahi_Fermi4FGL_2020}. This is done by employing a binned likelihood analysis to model and then subtract the contribution from point sources and extended sources of the 4FGL catalog located outside the masked regions.

To be consistent with this {\it modus operandi}, we also mask the simulated emission in the two \texttt{ROI}s following a similar method to that implemented in the LHAASO analysis. To do so, we use three different masks, referred to as $\mathcal{A}$, $\mathcal{B}$, and $\mathcal{A}\cup\mathcal{B}$ (see Appendix~\ref{app:mask} for additional details):
\begin{itemize}
    \item $\mathcal{A}$ removes the inner Galactic Plane ($l \leq 70^\circ$, $|b|\leq 1.5^\circ$) and the Local Arm (disk centered on [$l = 73.5^\circ$, $b=0^\circ$]).  This mask is designed to emulate the removal of Galactic sources, such as pulsar wind nebulae (PWNe) or supernova remnants (SNRs), which are not considered in our realization of the Milky Way and that are predominantly located along the Galactic Plane, within the first and fourth Galactic Quadrants. 
    \item $\mathcal{B}$ masks all those positions of the sky where the significance of the emission from YMSCs at 100 TeV, evaluated as described in Appendix \ref{app:mask}, is larger than 5 times the diffuse emission.    
    This mask is intended to remove all clusters whose emission should be bright enough to be detectable by LHAASO. We here assume that all the clusters detected by LHAASO should be also detected by Fermi-LAT. Hence, we do not mask nor remove the emission from YMSCs that should have been detected by Fermi-LAT but not by LHAASO. We will comment a posteriori the implications of this assumption.
    \item $\mathcal{A}\cup\mathcal{B}$ is the union of $\mathcal{A}$ and $\mathcal{B}$. The flux obtained using this mask is the most suitable for comparison with data, as $\mathcal{A}\cup\mathcal{B}$ most closely matches the masks used in the works by \cite{Cao_LHAASODiffuseGamma_2023} and \cite{Zhang_LHAASOandFermiLATDiffuseGamma_2023}.
\end{itemize}

Note that in \texttt{ROI2} the mask $\mathcal{A}$ is not defined, so in this region one will have $\mathcal{B} \equiv \mathcal{A}\cup\mathcal{B}$ (see Appendix~\ref{app:mask}). The bottom panels in Figure\,\ref{fig:SC_distrib} reports an example of the mask $\mathcal{A}\cup\mathcal{B}$ in both considered regions for a random realization of the Milky Way YMSCs population.

\subsection{Comparison with the observed galactic diffuse $\gamma$-ray emission}
\begin{figure*}[t]
\begin{center}
\includegraphics[width=0.95\textwidth]{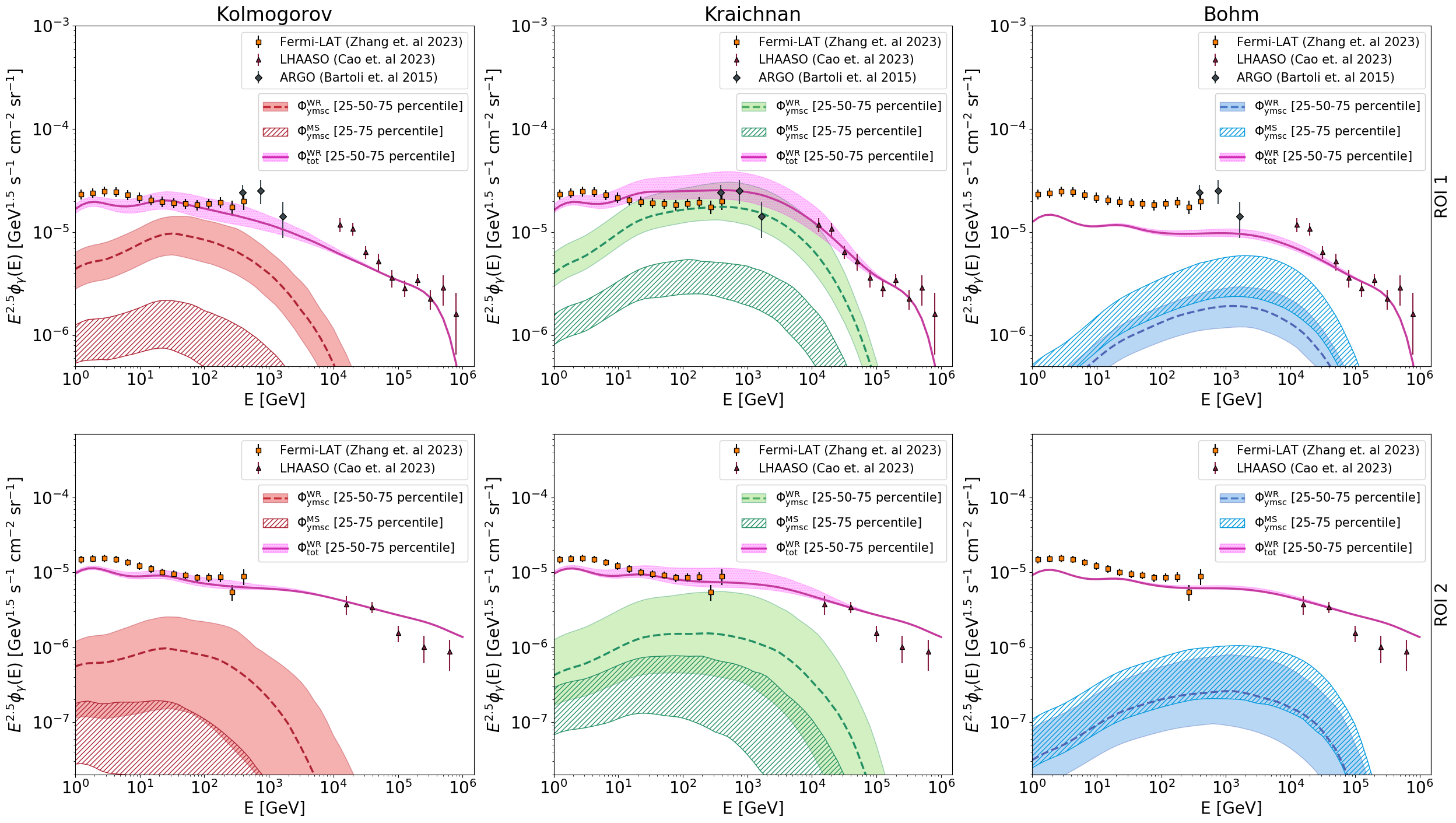} 
\caption{Contribution to the diffuse $\gamma$-ray emission from a synthetic population of YMSCs in \texttt{ROI1} (top row) and \texttt{ROI2} (bottom row) for the three different cases of diffusion coefficient considered: Kolmogorov (left column), Kraichnan (central column) and Bohm-like (right column). The spectrum is calculated after applying the mask $\mathcal{A}\cup\mathcal{B}$. In each plot, the dashed line represents the median flux per energy bin after considering 100 different realisations of the Milky Way population of YMSCs, while the associated shaded region encloses the 25-75 percentile flux. The striped region instead encompasses the 25-75 percentile flux when WR stars are not included in the stellar mock populations. The violet solid lines in the right column and their associated shaded regions show the median and the 25-75 percentile diffuse $\gamma$-ray flux after the addition of emission from the CR sea.}
\label{fig:ROI1and2}
\end{center}
\end{figure*}

To account for the uncertainty associated with the stochastic nature of the generation of the synthetic population of YMSCs, we simulate 100 different realizations of the Milky Way and calculate, for each of them, the contribution to the diffuse $\gamma$-ray flux, adopting different masks and assumptions on the particle diffusion. Figure~\ref{fig:ROI1and2} shows the YMSCs contribution to the diffuse $\gamma$-ray emission in \texttt{ROI1} (top row) and \texttt{ROI2} (bottom row) for the three types of diffusion coefficient considered. The emission from YMSCs is calculated after applying $\mathcal{A}\cup\mathcal{B}$ in both \texttt{ROI1} and \texttt{ROI2}.  In all plots the dashed lines represent the median flux per energy bin over the different (100) realisations of the Milky Way. The shaded areas cover instead the range between the 25 and 75 percentile of the distribution, including (color filled) or neglecting (striped) the contribution from WR stars.

In both \texttt{ROI}s, the $\gamma$-ray flux due to unresolved YMSCs is negligible when considering only the wind power from main sequence stars ($\phi_{\rm ymsc}^{\rm MS}$). On the contrary, when WR stars are added to the stellar populations ($\phi_{\rm ymsc}^{\rm WR}$) the contribution from YMSCs can account in some cases for a substantial fraction of the measured flux. To better visualize the relative importance of the emission from YMSCs, we also show the total expected diffuse $\gamma$-ray flux ($\phi_{\rm tot}^{\rm WR}$) when the contribution from the bulk of diffuse CRs (i.e. the CR \textit{sea}) is considered: $\phi_{\rm tot}^{\rm WR}=F_\gamma+\phi_{\rm ymsc}^{\rm WR}$, where $F_\gamma$ is the flux from the CR sea (see Appendix~\ref{app:CRseaFlux}).

More precisely, in \texttt{ROI1}, the contribution of YMSCs in the Kolmogorov case can explain the diffuse $\gamma$-ray emission from $\sim$10\,GeV to $\sim$1\,TeV. In the Kraichnan scenario, however, the contribution of YMSCs is higher and the total $\gamma$-ray emission exceeds that observed from $\sim$10\,GeV to $\sim$10\,TeV by $\sim$20\%. This implies that in the Kraichnan case, the contribution from YMSCs is able to totally account for the excess observed in the diffuse galactic emission if the product between $\epsilon_{\rm cr}$ and $n_0$ is decreased by a factor $\sim$0.8. On the other hand, in the Bohm case, the contribution from YMSCs to the diffuse emission is negligible. This is because, when Bohm diffusion is adopted, many more clusters have $\gamma$-ray emission above the LHAASO threshold and are hence removed by applying the mask $\mathcal{B}$ (see Appendix~\ref{app:mask}).

In \texttt{ROI2} the contribution from YMSCs is on average negligible, regardless of the presence of WR stars, with the exception of the Kraichnan case, where YMSCs could significantly contribute at $\sim$100\,GeV. However, no solid conclusion can be drawn due to the large statistical uncertainties in this region. In fact, due to the low content of YMSCs in the outer Galaxy, some realisations have a much larger $\gamma$-ray flux. However, it is likely that the knowledge of the population of YMSCs in the outer region of the Galaxy is complete or quasi-complete. Consequently, one could consider to rely directly on observed clusters from Gaia \cite[see, e.g.][]{Celli+2024, Mitchell+2024}, rather than using our approach.

These results must be considered with a few caveats in mind. First, the diffuse emission below $\sim$100\,GeV is likely to be slightly overestimated, as we did not subtract the emission from those YMSCs that would have been detected by Fermi-LAT but not by LHAASO. In any case, we do not expect this choice to lead to a large error because the most powerful clusters are subtracted anyway. Second, as mentioned in Section~\ref{sec:CR_model}, because of radiation losses, the size of the bubble could be smaller, reducing the $\gamma$-ray emission of each cluster. On average a radius reduction between 20\% and 50\% may be expected. If we assume that every bubble is systematically reduced by 20\% we found that the overall flux is reduced by roughly the same factor. On the other hand, for a reduction of 50\%, the flux is reduced by a factor $\sim 8$, and the emission from YMSCs is drastically suppressed. However, the latter case is somehow unrealistic for two reasons: first, radiation losses are in general expected to appear after $\sim$10\,Myr \citep{Silich_RadPressYMSCs_2013}, and second, one should also consider the re-heating of the gas induced by supernova explosion in the cluster, which can compensate for the energy losses. \ste{Finally, we note that adding the contribution of SNe the total predicted flux would likely overshoot the data for the current values of CR acceleration efficiency and gas density, especially in the Kraichnan scenario. We expect that consistency with the data could still be recovered by adjusting the product between $\epsilon_{\rm cr}$ and $n_0$, while still keeping both parameters in a reasonable range. However, fitting the observed spectrum is beyond the scope of this work, as it would also require the inclusion of additional populations of unresolved galactic $\gamma$-ray sources.}

To summarize, we assess that non-resolved YMSCs do provide a considerable contribution to the diffuse galactic $\gamma$-ray emission in the inner region of the Galactic plane if the particle diffusion coefficient follows a Kolmogorov-like or Kraichnan-like spectral energy dependence. Such a conclusion is strengthened by the fact that our estimate neglects the emission coming from particles accelerated at SNRs.

\section{Conclusions}
\label{sec:conclusions}
In recent years, the improving sensitivity of $\gamma$-ray telescopes has led to the identification of several YMSCs as $\gamma$-ray sources. Nonetheless, nowadays, the detection of individual YMSCs still remains a challenging task due to their low surface brightness. This implies that most of the $\gamma$-ray flux from YMSCs is likely detected as an unresolved contribution to the diffuse Galactic $\gamma$-ray emission. Indeed, recent analyses of the diffuse $\gamma$-ray emission have shown a prominent excess, likely related to unresolved sources, ranging from a few tens of GeV up to hundreds of TeV. In this work, we estimated a lower limit \ste{to} the contribution of YMSCs to the diffuse emission by simulating the entire Galactic populations of YMSCs, starting from observed properties of local clusters. We modeled the $\gamma$-ray spectrum of each synthetic YMSC considering a purely hadronic scenario and assuming that particle acceleration occurs at the collective cluster wind termination shock. We found that the contribution of the cluster population to the diffuse emission is sizeable above few tens of GeV in the inner region of the Milky Way when the particle transport in the wind-blown bubble is governed by Kolmogorov or Kraichnan type diffusion. 
In the outer region of the Milky Way we could not draw any solid conclusion due to the large statistical uncertainties related to the realisation of the YMSC population. A more reliable estimate in this region should be made by directly using existing YMSC surveys. 

The contribution of Wolf-Rayet stars turns out to dominate the collective wind power and hence the CR production, so that uncertainties in the modeling of these stars are the main caveat to our conclusions. Another source of uncertainty comes from the neglect of CRs accelerated by SNRs within the cluster. These are likely to further enhance the predicted $\gamma$-ray flux. However, at the moment, some important aspects concerning the evolution of SNRs inside a bubble and the associated particle acceleration are still unclear and require a dedicated effort to be fully understood.

All the effects we neglected, if properly taken into account, can only lead to estimate a larger contribution of YMSCs to the diffuse $\gamma$-ray background, possibly reinforcing our conclusions: YMSCs provide a non-negligible contribution to the diffuse $\gamma$-ray emission of the Galaxy, and their contribution must be taken into account in any future realistic model of the Galactic diffuse emission.

\begin{acknowledgements}
We thank Pr. Ruo-Yu Liu for providing the detector exposure on behalf of the LHAASO collaboration. This work was partially funded by the European Union – NextGenerationEU RRF M4C2 1.1 under grant PRIN-MUR 2022TJW4EJ
SM and GM are partially supported by the INAF Theory Grant 2022 {\it ‘‘Star Clusters As Cosmic Ray Factories''} and by the INAF Mini Grant 2023 {\it "Probing Young Massive Stellar Cluster as Cosmic Ray Factories"}.       
\end{acknowledgements}

%
%
\bibliographystyle{aa}
\bibliography{Bibliography}

\begin{thebibliography}{68}
\expandafter\ifx\csname natexlab\endcsname\relax\def\natexlab#1{#1}\fi

\bibitem[{{Abdollahi} {et~al.}(2020){Abdollahi}, {Acero}, {Ackermann},
  {Ajello}, {Atwood}, {Axelsson}, {Baldini}, {Ballet}, {Barbiellini},
  {Bastieri}, {Becerra Gonzalez}, {Bellazzini}, {Berretta}, {Bissaldi},
  {Blandford}, {Bloom}, {Bonino}, {Bottacini}, {Brandt}, {Bregeon}, {Bruel},
  {Buehler}, {Burnett}, {Buson}, {Cameron}, {Caputo}, {Caraveo}, {Casandjian},
  {Castro}, {Cavazzuti}, {Charles}, {Chaty}, {Chen}, {Cheung}, {Chiaro},
  {Ciprini}, {Cohen-Tanugi}, {Cominsky}, {Coronado-Bl{\'a}zquez}, {Costantin},
  {Cuoco}, {Cutini}, {D'Ammando}, {DeKlotz}, {de la Torre Luque}, {de Palma},
  {Desai}, {Digel}, {Di Lalla}, {Di Mauro}, {Di Venere}, {Dom{\'\i}nguez},
  {Dumora}, {Fana Dirirsa}, {Fegan}, {Ferrara}, {Franckowiak}, {Fukazawa},
  {Funk}, {Fusco}, {Gargano}, {Gasparrini}, {Giglietto}, {Giommi}, {Giordano},
  {Giroletti}, {Glanzman}, {Green}, {Grenier}, {Griffin}, {Grondin}, {Grove},
  {Guiriec}, {Harding}, {Hayashi}, {Hays}, {Hewitt}, {Horan},
  {J{\'o}hannesson}, {Johnson}, {Kamae}, {Kerr}, {Kocevski}, {Kovac'evic'},
  {Kuss}, {Landriu}, {Larsson}, {Latronico}, {Lemoine-Goumard}, {Li},
  {Liodakis}, {Longo}, {Loparco}, {Lott}, {Lovellette}, {Lubrano}, {Madejski},
  {Maldera}, {Malyshev}, {Manfreda}, {Marchesini}, {Marcotulli},
  {Mart{\'\i}-Devesa}, {Martin}, {Massaro}, {Mazziotta}, {McEnery}, {Mereu},
  {Meyer}, {Michelson}, {Mirabal}, {Mizuno}, {Monzani}, {Morselli},
  {Moskalenko}, {Negro}, {Nuss}, {Ojha}, {Omodei}, {Orienti}, {Orlando},
  {Ormes}, {Palatiello}, {Paliya}, {Paneque}, {Pei}, {Pe{\~n}a-Herazo},
  {Perkins}, {Persic}, {Pesce-Rollins}, {Petrosian}, {Petrov}, {Piron}, {Poon},
  {Porter}, {Principe}, {Rain{\`o}}, {Rando}, {Razzano}, {Razzaque}, {Reimer},
  {Reimer}, {Remy}, {Reposeur}, {Romani}, {Saz Parkinson}, {Schinzel},
  {Serini}, {Sgr{\`o}}, {Siskind}, {Smith}, {Spandre}, {Spinelli}, {Strong},
  {Suson}, {Tajima}, {Takahashi}, {Tak}, {Thayer}, {Thompson}, {Tibaldo},
  {Torres}, {Torresi}, {Valverde}, {Van Klaveren}, {van Zyl}, {Wood},
  {Yassine}, \& {Zaharijas}}]{Abdollahi_Fermi4FGL_2020}
{Abdollahi}, S., {Acero}, F., {Ackermann}, M., {et~al.} 2020, \apjs, 247, 33

\bibitem[{{Abeysekara} {et~al.}(2021){Abeysekara}, {Albert}, {Alfaro},
  {Alvarez}, {Camacho}, {Arteaga-Vel{\'a}zquez}, {Arunbabu}, {Rojas},
  {Solares}, {Baghmanyan}, {Belmont-Moreno}, {BenZvi}, {Blandford}, {Brisbois},
  {Caballero-Mora}, {Capistr{\'a}n}, {Carrami{\~n}ana}, {Casanova}, {Cotti},
  {Le{\'o}n}, {De la Fuente}, {Hernandez}, {Dingus}, {DuVernois}, {Durocher},
  {D{\'\i}az-V{\'e}lez}, {Ellsworth}, {Engel}, {Espinoza}, {Fan}, {Fang},
  {Fleischhack}, {Fraija}, {Galv{\'a}n-G{\'a}mez}, {Garcia},
  {Garc{\'\i}a-Gonz{\'a}lez}, {Garfias}, {Giacinti}, {Gonz{\'a}lez}, {Goodman},
  {Harding}, {Hernandez}, {Hinton}, {Hona}, {Huang}, {Hueyotl-Zahuantitla},
  {H{\"u}ntemeyer}, {Iriarte}, {Jardin-Blicq}, {Joshi}, {Kieda}, {Lara}, {Lee},
  {Vargas}, {Linnemann}, {Longinotti}, {Luis-Raya}, {Lundeen}, {Malone},
  {Martinez}, {Martinez-Castellanos}, {Mart{\'\i}nez-Castro}, {Matthews},
  {Miranda-Romagnoli}, {Morales-Soto}, {Moreno}, {Mostaf{\'a}}, {Nayerhoda},
  {Nellen}, {Newbold}, {Nisa}, {Noriega-Papaqui}, {Olivera-Nieto}, {Omodei},
  {Peisker}, {P{\'e}rez Araujo}, {P{\'e}rez-P{\'e}rez}, {Ren}, {Rho},
  {Rosa-Gonz{\'a}lez}, {Ruiz-Velasco}, {Salazar}, {Greus}, {Sandoval},
  {Schneider}, {Schoorlemmer}, {Serna}, {Smith}, {Springer}, {Surajbali},
  {Tollefson}, {Torres}, {Torres-Escobedo}, {Ure{\~n}a-Mena}, {Weisgarber},
  {Werner}, {Willox}, {Zepeda}, {Zhou}, {De Le{\'o}n}, \&
  {{\'A}lvarez}}]{Abeysekara_CygOB2HAWC_2021}
{Abeysekara}, A.~U., {Albert}, A., {Alfaro}, R., {et~al.} 2021, Nature
  Astronomy, 5, 465

\bibitem[{{Abramowski} {et~al.}(2012){Abramowski}, {Acero}, {Aharonian},
  {Akhperjanian}, {Anton}, {Balzer}, {Barnacka}, {Barres de Almeida},
  {Becherini}, {Becker}, {Behera}, {Bernl{\"o}hr}, {Birsin}, {Biteau},
  {Bochow}, {Boisson}, {Bolmont}, {Bordas}, {Brucker}, {Brun}, {Brun}, {Bulik},
  {B{\"u}sching}, {Carrigan}, {Casanova}, {Cerruti}, {Chadwick}, {Charbonnier},
  {Chaves}, {Cheesebrough}, {Chounet}, {Clapson}, {Coignet}, {Cologna},
  {Conrad}, {Dalton}, {Daniel}, {Davids}, {Degrange}, {Deil}, {Dickinson},
  {Djannati-Ata{\"\i}}, {Domainko}, {O'C. Drury}, {Dubois}, {Dubus}, {Dutson},
  {Dyks}, {Dyrda}, {Egberts}, {Eger}, {Espigat}, {Fallon}, {Farnier}, {Fegan},
  {Feinstein}, {Fernandes}, {Fiasson}, {Fontaine}, {F{\"o}rster},
  {F{\"u}{\ss}ling}, {Gallant}, {Gast}, {G{\'e}rard}, {Gerbig}, {Giebels},
  {Glicenstein}, {Gl{\"u}ck}, {Goret}, {G{\"o}ring}, {H{\"a}ffner}, {Hague},
  {Hampf}, {Hauser}, {Heinz}, {Heinzelmann}, {Henri}, {Hermann}, {Hinton},
  {Hoffmann}, {Hofmann}, {Hofverberg}, {Holler}, {Horns}, {Jacholkowska}, {de
  Jager}, {Jahn}, {Jamrozy}, {Jung}, {Kastendieck}, {Katarzy{\'n}ski}, {Katz},
  {Kaufmann}, {Keogh}, {Khangulyan}, {Kh{\'e}lifi}, {Klochkov}, {Klu{\.z}niak},
  {Kneiske}, {Komin}, {Kosack}, {Kossakowski}, {Laffon}, {Lamanna}, {Lennarz},
  {Lohse}, {Lopatin}, {Lu}, {Marandon}, {Marcowith}, {Masbou}, {Maurin},
  {Maxted}, {Mayer}, {McComb}, {Medina}, {M{\'e}hault}, {Moderski}, {Moulin},
  {Naumann}, {Naumann-Godo}, {de Naurois}, {Nedbal}, {Nekrassov}, {Nguyen},
  {Nicholas}, {Niemiec}, {Nolan}, {Ohm}, {de O{\~n}a Wilhelmi}, {Opitz},
  {Ostrowski}, {Oya}, {Panter}, {Paz Arribas}, {Pedaletti}, {Pelletier},
  {Petrucci}, {Pita}, {P{\"u}hlhofer}, {Punch}, {Quirrenbach}, {Raue},
  {Rayner}, {Reimer}, {Reimer}, {Renaud}, {de Los Reyes}, {Rieger}, {Ripken},
  {Rob}, {Rosier-Lees}, {Rowell}, {Rudak}, {Rulten}, {Ruppel}, {Sahakian},
  {Sanchez}, {Santangelo}, {Schlickeiser}, {Sch{\"o}ck}, {Schulz}, {Schwanke},
  {Schwarzburg}, {Schwemmer}, {Sheidaei}, {Sikora}, {Skilton}, {Sol},
  {Spengler}, {Stawarz}, {Steenkamp}, {Stegmann}, {Stinzing}, {Stycz},
  {Sushch}, {Szostek}, {Tavernet}, {Terrier}, {Tluczykont}, {Valerius}, {van
  Eldik}, {Vasileiadis}, {Venter}, {Vialle}, {Viana}, {Vincent}, {V{\"o}lk},
  {Volpe}, {Vorobiov}, {Vorster}, {Wagner}, {Ward}, {White}, {Wierzcholska},
  {Zacharias}, {Zajczyk}, {Zdziarski}, {Zech}, \&
  {Zechlin}}]{Abramowski_Wd1VHE_2012}
{Abramowski}, A., {Acero}, F., {Aharonian}, F., {et~al.} 2012, \aap, 537, A114

\bibitem[{{Aharonian} {et~al.}(2022){Aharonian}, {Ashkar}, {Backes}, {Barbosa
  Martins}, {Becherini}, {Berge}, {Bi}, {B{\"o}ttcher}, {de Bony de Lavergne},
  {Bradascio}, {Brose}, {Brun}, {Bulik}, {Burger-Scheidlin}, {Cangemi},
  {Caroff}, {Casanova}, {Cerruti}, {Chand}, {Chandra}, {Chen}, {Chibueze},
  {Cristofari}, {Damascene Mbarubucyeye}, {Djannati-Ata{\"\i}}, {Ernenwein},
  {Feijen}, {Fichet de Clairfontaine}, {Fontaine}, {Funk}, {Gabici}, {Gallant},
  {Ghafourizadeh}, {Giavitto}, {Giunti}, {Glawion}, {Glicenstein}, {Goswami},
  {Grondin}, {H{\"a}rer}, {Haupt}, {Hinton}, {H{\"o}rbe}, {Hofmann}, {Holch},
  {Holler}, {Horns}, {Jamrozy}, {Joshi}, {Jung-Richardt}, {Kasai},
  {Katarzy{\'n}ski}, {Katz}, {Kh{\'e}lifi}, {Klu{\'z}niak}, {Komin}, {Kosack},
  {Kostunin}, {Kukec Mezek}, {Lang}, {Le Stum}, {Lemi{\`e}re},
  {Lemoine-Goumard}, {Lenain}, {Leuschner}, {Lohse}, {Luashvili}, {Lypova},
  {Mackey}, {Majumdar}, {Malyshev}, {Marandon}, {Marchegiani}, {Marcowith},
  {Mart{\'\i}-Devesa}, {Marx}, {Maurin}, {Meyer}, {Mitchell}, {Moderski},
  {Mohrmann}, {Montanari}, {Moulin}, {Muller}, {Murach}, {Nakashima}, {de
  Naurois}, {Nayerhoda}, {Niemiec}, {Ohm}, {Olivera-Nieto}, {de Ona Wilhelmi},
  {Ostrowski}, {Panny}, {Panter}, {Parsons}, {Peron}, {Prokhorov},
  {P{\"u}hlhofer}, {Punch}, {Quirrenbach}, {Rauth}, {Reichherzer}, {Reimer},
  {Reimer}, {Renaud}, {Reville}, {Rieger}, {Rowell}, {Rudak}, {Ruiz-Velasco},
  {Sahakian}, {Salzmann}, {Sanchez}, {Santangelo}, {Sasaki}, {Sch{\"u}ssler},
  {Schutte}, {Schwanke}, {Shapopi}, {Specovius}, {Spencer}, {Stawarz},
  {Steenkamp}, {Steinmassl}, {Steppa}, {Sushch}, {Suzuki}, {Takahashi},
  {Tanaka}, {Terrier}, {Thorpe-Morgan}, {Tsirou}, {Tsuji}, {Tuffs}, {Unbehaun},
  {van Eldik}, {van Soelen}, {Vecchi}, {Veh}, {Venter}, {Vink}, {Wagner},
  {White}, {Wierzcholska}, {Wun Wong}, {Zacharias}, {Zargaryan}, {Zdziarski},
  {Zhu}, {Zouari}, {{\.Z}ywucka}, {Blackwell}, {Braiding}, {Burton}, {Cubuk},
  {Filipovi{\'c}}, {Tothill}, \& {Wong}}]{Aharonian_HESSWesterlund1_2022}
{Aharonian}, F., {Ashkar}, H., {Backes}, M., {et~al.} 2022, arXiv e-prints,
  arXiv:2207.10921

\bibitem[{{Aharonian} {et~al.}(2019){Aharonian}, {Yang}, \& {de O{\~n}a
  Wilhelmi}}]{Aharonian_MSCs_2019}
{Aharonian}, F., {Yang}, R., \& {de O{\~n}a Wilhelmi}, E. 2019, Nature
  Astronomy, 3, 561

\bibitem[{{Astiasarain} {et~al.}(2023){Astiasarain}, {Tibaldo.}, {Martin},
  {Kn{\"o}dlseder}, \& {Remy}}]{Astiasarain_CygCocoon_2023}
{Astiasarain}, X., {Tibaldo.}, L., {Martin}, P., {Kn{\"o}dlseder}, J., \&
  {Remy}, Q. 2023, arXiv e-prints, arXiv:2301.04504

\bibitem[{{Bartoli} {et~al.}(2014){Bartoli}, {Bernardini}, {Bi}, {Branchini},
  {Budano}, {Camarri}, {Cao}, {Cardarelli}, {Catalanotti}, {Chen}, {Chen},
  {Creti}, {Cui}, {Dai}, {D'Amone}, {Danzengluobu}, {De Mitri}, {D'Ettorre
  Piazzoli}, {Di Girolamo}, {Di Sciascio}, {Feng}, {Feng}, {Feng}, {Gou},
  {Guo}, {He}, {Hu}, {Hu}, {Iacovacci}, {Iuppa}, {Jia}, {Labaciren}, {Li},
  {Liguori}, {Liu}, {Liu}, {Liu}, {Lu}, {Ma}, {Ma}, {Mancarella}, {Mari},
  {Marsella}, {Martello}, {Mastroianni}, {Montini}, {Ning}, {Panareo},
  {Perrone}, {Pistilli}, {Ruggieri}, {Salvini}, {Santonico}, {Shen}, {Sheng},
  {Shi}, {Surdo}, {Tan}, {Vallania}, {Vernetto}, {Vigorito}, {Wang}, {Wu},
  {Wu}, {Xue}, {Yang}, {Yang}, {Yao}, {Yuan}, {Zha}, {Zhang}, {Zhang}, {Zhang},
  {Zhang}, {Zhao}, {Zhaxiciren}, {Zhaxisangzhu}, {Zhou}, {Zhu}, {Zhu}, {Zizzi},
  \& {ARGO-YBJ Collaboration}}]{Bartoli_CygOB2Argo_2014}
{Bartoli}, B., {Bernardini}, P., {Bi}, X.~J., {et~al.} 2014, \apj, 790, 152

\bibitem[{{Bartoli} {et~al.}(2015){Bartoli}, {Bernardini}, {Bi}, {Branchini},
  {Budano}, {Camarri}, {Cao}, {Cardarelli}, {Catalanotti}, {Chen}, {Chen},
  {Creti}, {Cui}, {Dai}, {D'Amone}, {Danzengluobu}, {De Mitri}, {D'Ettorre
  Piazzoli}, {Di Girolamo}, {Di Sciascio}, {Feng}, {Feng}, {Feng}, {Gou},
  {Guo}, {He}, {Hu}, {Hu}, {Iacovacci}, {Iuppa}, {Jia}, {Labaciren}, {Li},
  {Liguori}, {Liu}, {Liu}, {Liu}, {Lu}, {Ma}, {Ma}, {Mancarella}, {Mari},
  {Marsella}, {Martello}, {Mastroianni}, {Montini}, {Ning}, {Panareo},
  {Perrone}, {Pistilli}, {Ruggieri}, {Salvini}, {Santonico}, {Shen}, {Sheng},
  {Shi}, {Surdo}, {Tan}, {Vallania}, {Vernetto}, {Vigorito}, {Wang}, {Wu},
  {Wu}, {Xue}, {Yang}, {Yang}, {Yao}, {Yuan}, {Zha}, {Zhang}, {Zhang}, {Zhang},
  {Zhang}, {Zhao}, {Zhaxiciren}, {Zhaxisangzhu}, {Zhou}, {Zhu}, {Zhu}, {Zizzi},
  \& {ARGO-YBJ Collaboration}}]{Bartoli_ArgoDiffuse_2015}
{Bartoli}, B., {Bernardini}, P., {Bi}, X.~J., {et~al.} 2015, \apj, 806, 20

\bibitem[{{Blasi} \& {Morlino}(2024)}]{Blasi-Morlino:2024}
{Blasi}, P. \& {Morlino}, G. 2024, \mnras, 533, 561

\bibitem[{{Bonatto} \& {Bica}(2011)}]{Bonatto_SFRinSC_2011}
{Bonatto}, C. \& {Bica}, E. 2011, \mnras, 415, 2827

\bibitem[{{Buzzoni}(2002)}]{Buzzoni_TOtime_2002}
{Buzzoni}, A. 2002, \aj, 123, 1188

\bibitem[{{Bykov} {et~al.}(2020){Bykov}, {Marcowith}, {Amato}, {Kalyashova},
  {Kruijssen}, \& {Waxman}}]{Bykov_MSCTurbAcc_2020}
{Bykov}, A.~M., {Marcowith}, A., {Amato}, E., {et~al.} 2020, \ssr, 216, 42

\bibitem[{{Cantat-Gaudin} {et~al.}(2020){Cantat-Gaudin}, {Anders},
  {Castro-Ginard}, {Jordi}, {Romero-G{\'o}mez}, {Soubiran}, {Casamiquela},
  {Tarricq}, {Moitinho}, {Vallenari}, {Bragaglia}, {Krone-Martins}, \&
  {Kounkel}}]{Cantat-Gaudin:2020}
{Cantat-Gaudin}, T., {Anders}, F., {Castro-Ginard}, A., {et~al.} 2020, \aap,
  640, A1

\bibitem[{{Cao} {et~al.}(2023{\natexlab{a}}){Cao}, {Aharonian}, {An},
  {Axikegu}, {Bai}, {Bao}, {Bastieri}, {Bi}, {Bi}, {Cai}, {Cao}, {Cao}, {Cao},
  {Chang}, {Chang}, {Chen}, {Chen}, {Chen}, {Chen}, {Chen}, {Chen}, {Chen},
  {Chen}, {Chen}, {Chen}, {Chen}, {Chen}, {Cheng}, {Cheng}, {Cui}, {Cui},
  {Cui}, {Cui}, {Dai}, {Dai}, {Dai}, {Danzengluobu}, {della Volpe}, {Dong},
  {Duan}, {Fan}, {Fan}, {Fang}, {Fang}, {Feng}, {Feng}, {Feng}, {Feng}, {Feng},
  {Gabici}, {Gao}, {Gao}, {Gao}, {Gao}, {Gao}, {Gao}, {Ge}, {Geng}, {Giacinti},
  {Gong}, {Gou}, {Gu}, {Guo}, {Guo}, {Guo}, {Guo}, {Han}, {He}, {He}, {He},
  {He}, {He}, {Heller}, {Hor}, {Hou}, {Hou}, {Hou}, {Hu}, {Hu}, {Hu}, {Huang},
  {Huang}, {Huang}, {Huang}, {Huang}, {Huang}, {Huang}, {Ji}, {Jia}, {Jia},
  {Jiang}, {Jiang}, {Jiang}, {Jin}, {Kang}, {Ke}, {Kuleshov}, {Kurinov}, {Li},
  {Li}, {Li}, {Li}, {Li}, {Li}, {Li}, {Li}, {Li}, {Li}, {Li}, {Li}, {Li}, {Li},
  {Li}, {Li}, {Li}, {Li}, {Li}, {Liang}, {Liang}, {Lin}, {Liu}, {Liu}, {Liu},
  {Liu}, {Liu}, {Liu}, {Liu}, {Liu}, {Liu}, {Liu}, {Liu}, {Liu}, {Liu}, {Liu},
  {Lu}, {Luo}, {Lv}, {Ma}, {Ma}, {Ma}, {Mao}, {Min}, {Mitthumsiri}, {Mu},
  {Nan}, {Neronov}, {Ou}, {Pang}, {Pattarakijwanich}, {Pei}, {Qi}, {Qi},
  {Qiao}, {Qin}, {Ruffolo}, {Saiz}, {Semikoz}, {Shao}, {Shao}, {Shchegolev},
  {Sheng}, {Shu}, {Song}, {Stenkin}, {Stepanov}, {Su}, {Sun}, {Sun}, {Sun},
  {Tam}, {Tang}, {Tang}, {Tian}, {Wang}, {Wang}, {Wang}, {Wang}, {Wang},
  {Wang}, {Wang}, {Wang}, {Wang}, {Wang}, {Wang}, {Wang}, {Wang}, {Wang},
  {Wang}, {Wang}, {Wang}, {Wang}, {Wang}, {Wang}, {Wang}, {Wei}, {Wei}, {Wei},
  {Wen}, {Wu}, {Wu}, {Wu}, {Wu}, {Wu}, {Xi}, {Xia}, {Xia}, {Xiang}, {Xiao},
  {Xiao}, {Xin}, {Xin}, {Xing}, {Xiong}, {Xu}, {Xu}, {Xu}, {Xu}, {Xue}, {Yan},
  {Yan}, {Yan}, {Yang}, {Yang}, {Yang}, {Yang}, {Yang}, {Yang}, {Yang}, {Yang},
  {Yang}, {Yao}, {Yao}, {Ye}, {Yin}, {Yin}, {You}, {You}, {Yu}, {Yuan}, {Yue},
  {Zeng}, {Zeng}, {Zeng}, {Zha}, {Zhang}, {Zhang}, {Zhang}, {Zhang}, {Zhang},
  {Zhang}, {Zhang}, {Zhang}, {Zhang}, {Zhang}, {Zhang}, {Zhang}, {Zhang},
  {Zhang}, {Zhang}, {Zhang}, {Zhang}, {Zhang}, {Zhao}, {Zhao}, {Zhao}, {Zhao},
  {Zhao}, {Zheng}, {Zhou}, {Zhou}, {Zhou}, {Zhou}, {Zhou}, {Zhou}, {Zhou},
  {Zhu}, {Zhu}, {Zhu}, {Zhu}, \& {Zuo}}]{Cao_LHAASODiffuseGamma_2023}
{Cao}, Z., {Aharonian}, F., {An}, Q., {et~al.} 2023{\natexlab{a}}, arXiv
  e-prints, arXiv:2305.05372

\bibitem[{{Cao} {et~al.}(2023{\natexlab{b}}){Cao}, {Aharonian}, {An},
  {Axikegu}, {Bai}, {Bao}, {Bastieri}, {Bi}, {Bi}, {Cai}, {Cao}, {Cao}, {Cao},
  {Chang}, {Chang}, {Chen}, {Chen}, {Chen}, {Chen}, {Chen}, {Chen}, {Chen},
  {Chen}, {Chen}, {Chen}, {Chen}, {Chen}, {Cheng}, {Cheng}, {Cui}, {Cui},
  {Cui}, {Cui}, {Dai}, {Dai}, {Dai}, {Danzengluobu}, {della Volpe}, {Dong},
  {Duan}, {Fan}, {Fan}, {Fang}, {Fang}, {Feng}, {Feng}, {Feng}, {Feng}, {Feng},
  {Gabici}, {Gao}, {Gao}, {Gao}, {Gao}, {Gao}, {Gao}, {Ge}, {Geng}, {Giacinti},
  {Gong}, {Gou}, {Gu}, {Guo}, {Guo}, {Guo}, {Guo}, {Han}, {He}, {He}, {He},
  {He}, {He}, {Heller}, {Hor}, {Hou}, {Hou}, {Hou}, {Hu}, {Hu}, {Hu}, {Huang},
  {Huang}, {Huang}, {Huang}, {Huang}, {Huang}, {Huang}, {Ji}, {Jia}, {Jia},
  {Jiang}, {Jiang}, {Jiang}, {Jin}, {Kang}, {Ke}, {Kuleshov}, {Kurinov}, {Li},
  {Li}, {Li}, {Li}, {Li}, {Li}, {Li}, {Li}, {Li}, {Li}, {Li}, {Li}, {Li}, {Li},
  {Li}, {Li}, {Li}, {Li}, {Li}, {Liang}, {Liang}, {Lin}, {Liu}, {Liu}, {Liu},
  {Liu}, {Liu}, {Liu}, {Liu}, {Liu}, {Liu}, {Liu}, {Liu}, {Liu}, {Liu}, {Liu},
  {Lu}, {Luo}, {Lv}, {Ma}, {Ma}, {Ma}, {Mao}, {Min}, {Mitthumsiri}, {Mu},
  {Nan}, {Neronov}, {Ou}, {Pang}, {Pattarakijwanich}, {Pei}, {Qi}, {Qi},
  {Qiao}, {Qin}, {Ruffolo}, {S{\'a}iz}, {Semikoz}, {Shao}, {Shao},
  {Shchegolev}, {Sheng}, {Shu}, {Song}, {Stenkin}, {Stepanov}, {Su}, {Sun},
  {Sun}, {Sun}, {Tam}, {Tang}, {Tang}, {Tian}, {Wang}, {Wang}, {Wang}, {Wang},
  {Wang}, {Wang}, {Wang}, {Wang}, {Wang}, {Wang}, {Wang}, {Wang}, {Wang},
  {Wang}, {Wang}, {Wang}, {Wang}, {Wang}, {Wang}, {Wang}, {Wang}, {Wei}, {Wei},
  {Wei}, {Wen}, {Wu}, {Wu}, {Wu}, {Wu}, {Wu}, {Xi}, {Xia}, {Xia}, {Xiang},
  {Xiao}, {Xiao}, {Xin}, {Xin}, {Xing}, {Xiong}, {Xu}, {Xu}, {Xu}, {Xu}, {Xue},
  {Yan}, {Yan}, {Yan}, {Yang}, {Yang}, {Yang}, {Yang}, {Yang}, {Yang}, {Yang},
  {Yang}, {Yang}, {Yao}, {Yao}, {Ye}, {Yin}, {Yin}, {You}, {You}, {Yu}, {Yuan},
  {Yue}, {Zeng}, {Zeng}, {Zeng}, {Zha}, {Zhang}, {Zhang}, {Zhang}, {Zhang},
  {Zhang}, {Zhang}, {Zhang}, {Zhang}, {Zhang}, {Zhang}, {Zhang}, {Zhang},
  {Zhang}, {Zhang}, {Zhang}, {Zhang}, {Zhang}, {Zhang}, {Zhao}, {Zhao}, {Zhao},
  {Zhao}, {Zhao}, {Zheng}, {Zhou}, {Zhou}, {Zhou}, {Zhou}, {Zhou}, {Zhou},
  {Zhou}, {Zhu}, {Zhu}, {Zhu}, {Zhu}, \& {Zuo.}}]{Cao_LHAASOCat_2023}
{Cao}, Z., {Aharonian}, F., {An}, Q., {et~al.} 2023{\natexlab{b}}, arXiv
  e-prints, arXiv:2305.17030

\bibitem[{{Cao} {et~al.}(2019){Cao}, {della Volpe}, {Liu}, {Editors}, {:},
  {Bi}, {Chen}, {D'Ettorre Piazzoli}, {Feng}, {Jia}, {Li}, {Ma}, {Wang},
  {Zhang}, {Referees}, {:}, {Qie}, {Hu}, {Referees}, {:}, {S{\'a}iz}, {Yang},
  {Contributors}, {:}, {Addazi}, {Belotsky}, {Beylin}, {Bi}, {Che}, {Chen},
  {Cheng}, {Chiavassa}, {Cirelli}, {Di Sciascio}, {Esmaili}, {Fang},
  {Fornengo}, {Gou}, {Guo}, {Gan}, {Gong}, {Gu}, {He}, {He}, {Hou}, {Huang},
  {Huang}, {Kachekriess}, {Khlopov}, {Korchagin}, {Korochkin}, {Kuksa},
  {Ksenofontov}, {Liu}, {Liu}, {Liu}, {Marciano}, {Martineau-Huynh},
  {Martraire}, {Ma}, {Neronov}, {Panci}, {Pasechnick}, {Ruffolo}, {Sakharov},
  {Sala}, {Semikoz}, {Shchegolev}, {Serpico}, {Sheng}, {Stenkin}, {Tam},
  {Vernetto}, {Vallania}, {Volchanskiy}, {Wang}, {Wang}, {Wang}, {Wu}, {Wu},
  {Wu}, {Xiao}, {Yang}, {Yan}, {Yao}, {Yin}, {Yuan}, {Zhang}, {Zeng}, {Zhang},
  {Zhang}, {Zhou}, {Zhu}, \& {Zuo}}]{Cao_LHAASOScienceBook_2019}
{Cao}, Z., {della Volpe}, D., {Liu}, S., {et~al.} 2019, arXiv e-prints,
  arXiv:1905.02773

\bibitem[{{Cao} {et~al.}(2023{\natexlab{c}}){Cao}, {Li}, {Gau}, {Liu}, \&
  {Yang}}]{LHAASO_UHEgammaSuperPeV_2023}
{Cao}, Z., {Li}, C., {Gau}, C.~D., {Liu}, R.~Y., \& {Yang}, R.~Z.
  2023{\natexlab{c}}, arXiv e-prints, arXiv:2310.10100

\bibitem[{{Carroll} \& {Ostlie}(1996)}]{Carroll_IntroModAstro_1996}
{Carroll}, B.~W. \& {Ostlie}, D.~A. 1996, {An Introduction to Modern
  Astrophysics}

\bibitem[{{Celli} {et~al.}(2023){Celli}, {Specovius}, {Menchiari}, {Mitchell},
  \& {Morlino}}]{Celli+2024}
{Celli}, S., {Specovius}, A., {Menchiari}, S., {Mitchell}, A., \& {Morlino}, G.
  2023, arXiv e-prints, arXiv:2311.09089

\bibitem[{{Cesarsky} \& {Montmerle}(1983)}]{Cesarsky_GammaRayStellarWinds_1983}
{Cesarsky}, C.~J. \& {Montmerle}, T. 1983, \ssr, 36, 173

\bibitem[{{Churchwell} {et~al.}(2009){Churchwell}, {Babler}, {Meade},
  {Whitney}, {Benjamin}, {Indebetouw}, {Cyganowski}, {Robitaille}, {Povich},
  {Watson}, \& {Bracker}}]{Churchwell_GalBarGlimpse_2009}
{Churchwell}, E., {Babler}, B.~L., {Meade}, M.~R., {et~al.} 2009, \pasp, 121,
  213

\bibitem[{{Crowther}(2007)}]{Crowther_WRsReview_2007}
{Crowther}, P.~A. 2007, \araa, 45, 177

\bibitem[{{Crowther} {et~al.}(2010){Crowther}, {Schnurr}, {Hirschi}, {Yusof},
  {Parker}, {Goodwin}, \& {Kassim}}]{Crowther+2010}
{Crowther}, P.~A., {Schnurr}, O., {Hirschi}, R., {et~al.} 2010, \mnras, 408,
  731

\bibitem[{{Demircan} \& {Kahraman}(1991)}]{Demircan_StarsMRR_1991}
{Demircan}, O. \& {Kahraman}, G. 1991, \apss, 181, 313

\bibitem[{{Eldridge} \& {Vink}(2006)}]{Eldridge_WRMinMass_2006}
{Eldridge}, J.~J. \& {Vink}, J.~S. 2006, \aap, 452, 295

\bibitem[{{Faucher-Gigu{\`e}re} \&
  {Kaspi}(2006)}]{Faucher_IsolatedRadioPulsar_2006}
{Faucher-Gigu{\`e}re}, C.-A. \& {Kaspi}, V.~M. 2006, \apj, 643, 332

\bibitem[{{Gies}(1987)}]{Gies_ORunawayStars_1987}
{Gies}, D.~R. 1987, \apjs, 64, 545

\bibitem[{{Green} {et~al.}(2011){Green}, {Caswell}, {McClure-Griffiths},
  {Avison}, {Breen}, {Burton}, {Ellingsen}, {Fuller}, {Gray}, {Pestalozzi},
  {Thompson}, \& {Voronkov}}]{Green_NF3kpcArms_2011}
{Green}, J.~A., {Caswell}, J.~L., {McClure-Griffiths}, N.~M., {et~al.} 2011,
  \apj, 733, 27

\bibitem[{{H.~E.~S.~S. Collaboration} {et~al.}(2011){H.~E.~S.~S.
  Collaboration}, {Abramowski}, {Acero}, {Aharonian}, {Akhperjanian}, {Anton},
  {Barnacka}, {Barres de Almeida}, {Bazer-Bachi}, {Becherini}, {Becker},
  {Behera}, {Bernl{\"o}hr}, {Bochow}, {Boisson}, {Bolmont}, {Bordas}, {Borrel},
  {Brucker}, {Brun}, {Brun}, {Bulik}, {B{\"u}sching}, {Boutelier}, {Casanova},
  {Cerruti}, {Chadwick}, {Charbonnier}, {Chaves}, {Cheesebrough}, {Conrad},
  {Chounet}, {Clapson}, {Coignet}, {Dalton}, {Daniel}, {Davids}, {Degrange},
  {Deil}, {Dickinson}, {Djannati-Ata{\"\i}}, {Domainko}, {Drury}, {Dubois},
  {Dubus}, {Dyks}, {Dyrda}, {Egberts}, {Eger}, {Espigat}, {Fallon}, {Farnier},
  {Fegan}, {Feinstein}, {Fernandes}, {Fiasson}, {F{\"o}rster}, {Fontaine},
  {F{\"u}{\ss}ling}, {Gabici}, {Gallant}, {G{\'e}rard}, {Gerbig}, {Giebels},
  {Glicenstein}, {Gl{\"u}ck}, {Goret}, {G{\"o}ring}, {Hague}, {Hampf},
  {Hauser}, {Heinz}, {Heinzelmann}, {Henri}, {Hermann}, {Hinton}, {Hoffmann},
  {Hofmann}, {Hofverberg}, {Holleran}, {Hoppe}, {Horns}, {Jacholkowska}, {de
  Jager}, {Jahn}, {Jung}, {Katarzy{\'n}ski}, {Katz}, {Kaufmann}, {Kerschhaggl},
  {Khangulyan}, {Kh{\'e}lifi}, {Keogh}, {Klochkov}, {Klu{\'z}niak}, {Kneiske},
  {Komin}, {Kosack}, {Kossakowski}, {Lamanna}, {Lenain}, {Lennarz}, {Lohse},
  {Lu}, {Marandon}, {Marcowith}, {Masbou}, {Maurin}, {McComb}, {Medina},
  {M{\'e}hault}, {Moderski}, {Moulin}, {Naumann-Godo}, {de Naurois}, {Nedbal},
  {Nekrassov}, {Nguyen}, {Nicholas}, {Niemiec}, {Nolan}, {Ohm}, {Olive}, {de
  O{\~n}a Wilhelmi}, {Opitz}, {Orford}, {Ostrowski}, {Panter}, {Paz Arribas},
  {Pedaletti}, {Pelletier}, {Petrucci}, {Pita}, {P{\"u}hlhofer}, {Punch},
  {Quirrenbach}, {Raubenheimer}, {Raue}, {Rayner}, {Reimer}, {Reimer},
  {Renaud}, {de los Reyes}, {Rieger}, {Ripken}, {Rob}, {Rosier-Lees}, {Rowell},
  {Rudak}, {Rulten}, {Ruppel}, {Ryde}, {Sahakian}, {Santangelo},
  {Schlickeiser}, {Sch{\"o}ck}, {Sch{\"o}nwald}, {Schwanke}, {Schwarzburg},
  {Schwemmer}, {Shalchi}, {Sushch}, {Sikora}, {Skilton}, {Sol}, {Spengler},
  {Stawarz}, {Steenkamp}, {Stegmann}, {Stinzing}, {Szostek}, {Tam}, {Tavernet},
  {Terrier}, {Tibolla}, {Tluczykont}, {Valerius}, {van Eldik}, {Vasileiadis},
  {Venter}, {Vialle}, {Vincent}, {Vivier}, {V{\"o}lk}, {Volpe}, {Wagner},
  {Ward}, {Zdziarski}, {Zech}, {Zechlin}, {Fukui}, {Furukawa}, {Ohama}, {Sano},
  {Dawson}, {Kawamura}, \& {H.~E.~S.~S. Collaboration}}]{HESS_Wd2_2011}
{H.~E.~S.~S. Collaboration}, {Abramowski}, A., {Acero}, F., {et~al.} 2011,
  \aap, 525, A46

\bibitem[{{Hou} \& {Han}(2014)}]{Hou_MWStructure_2014}
{Hou}, L.~G. \& {Han}, J.~L. 2014, \aap, 569, A125

\bibitem[{Kafexhiu {et~al.}(2014)Kafexhiu, Aharonian, Taylor, \&
  Vila}]{Kafexhiu_SigmaPi0Gamma_2014}
Kafexhiu, E., Aharonian, F., Taylor, A.~M., \& Vila, G.~S. 2014, Physical
  Review D, 90, 123014, arXiv: 1406.7369

\bibitem[{{Klepach} {et~al.}(2000){Klepach}, {Ptuskin}, \&
  {Zirakashvili}}]{Klepach_WindWindInteractionCR_2000}
{Klepach}, E.~G., {Ptuskin}, V.~S., \& {Zirakashvili}, V.~N. 2000,
  Astroparticle Physics, 13, 161

\bibitem[{{Kroupa}(2001)}]{Kroupa_IMF_2001}
{Kroupa}, P. 2001, \mnras, 322, 231

\bibitem[{{Kudritzki} \& {Puls}(2000)}]{Kudritzki_WindsHotStars_2000}
{Kudritzki}, R.-P. \& {Puls}, J. 2000, \araa, 38, 613

\bibitem[{{Lada} \& {Lada}(2003)}]{Lada_SCReview_2003}
{Lada}, C.~J. \& {Lada}, E.~A. 2003, \araa, 41, 57

\bibitem[{{Lancaster} {et~al.}(2021){Lancaster}, {Ostriker}, {Kim}, \&
  {Kim}}]{Lancaster_StellarWindCooledII_2021}
{Lancaster}, L., {Ostriker}, E.~C., {Kim}, J.-G., \& {Kim}, C.-G. 2021, \apj,
  914, 90

\bibitem[{{Lipari} \& {Vernetto}(2018)}]{Lipari2018}
{Lipari}, P. \& {Vernetto}, S. 2018, \prd, 98, 043003

\bibitem[{{Lipari} \& {Vernetto}(2020)}]{Lipari2019}
{Lipari}, P. \& {Vernetto}, S. 2020, Astroparticle Physics, 120, 102441

\bibitem[{{Liu} {et~al.}(2022){Liu}, {Yang}, \& {Chen}}]{Liu2022M17}
{Liu}, B., {Yang}, R.-z., \& {Chen}, Z. 2022, \mnras, 513, 4747

\bibitem[{{Menchiari}(2023)}]{Menchiari_ProbingSCsCRsPhD_2023}
{Menchiari}, S. 2023, arXiv e-prints, arXiv:2307.03477

\bibitem[{{Menchiari} {et~al.}(2024){Menchiari}, {Morlino}, {Amato},
  {Bucciantini}, \& {Beltr{\'a}n}}]{Menchiari_CygOB2_2024}
{Menchiari}, S., {Morlino}, G., {Amato}, E., {Bucciantini}, N., \&
  {Beltr{\'a}n}, M.~T. 2024, \aap, 686, A242

\bibitem[{{Mitchell} {et~al.}(2024){Mitchell}, {Morlino}, {Celli}, {Menchiari},
  \& {Specovius}}]{Mitchell+2024}
{Mitchell}, A. M.~W., {Morlino}, G., {Celli}, S., {Menchiari}, S., \&
  {Specovius}, A. 2024, arXiv e-prints, arXiv:2403.16650

\bibitem[{{Mori}(2009)}]{Mori2009}
{Mori}, M. 2009, Astroparticle Physics, 31, 341

\bibitem[{{Morlino} {et~al.}(2021){Morlino}, {Blasi}, {Peretti}, \&
  {Cristofari}}]{Morlino_2021}
{Morlino}, G., {Blasi}, P., {Peretti}, E., \& {Cristofari}, P. 2021, \mnras,
  504, 6096

\bibitem[{{Niedzielski} \& {Skorzynski}(2002)}]{Niedzielski_2002}
{Niedzielski}, A. \& {Skorzynski}, W. 2002, \actaa, 52, 81

\bibitem[{{Nieuwenhuijzen} \& {de Jager}(1990)}]{Nieuwenhuijzen_Mdot_1990}
{Nieuwenhuijzen}, H. \& {de Jager}, C. 1990, \aap, 231, 134

\bibitem[{{Nugis} \& {Lamers}(2000)}]{Nugis_MdotWR_2000}
{Nugis}, T. \& {Lamers}, H.~J.~G.~L.~M. 2000, \aap, 360, 227

\bibitem[{{Parker} \& {Goodwin}(2007{\natexlab{a}})}]{Parker-Goodwin:2007}
{Parker}, R.~J. \& {Goodwin}, S.~P. 2007{\natexlab{a}}, \mnras, 380, 1271

\bibitem[{{Parker} \&
  {Goodwin}(2007{\natexlab{b}})}]{Parker_OstarClusters_2007}
{Parker}, R.~J. \& {Goodwin}, S.~P. 2007{\natexlab{b}}, \mnras, 380, 1271

\bibitem[{{Peron} \& {Aharonian}(2022)}]{Peron2022}
{Peron}, G. \& {Aharonian}, F. 2022, \aap, 659, A57

\bibitem[{Peron {et~al.}(2024)Peron, Casanova, Gabici, Baghmanyan, \&
  Aharonian}]{Peron_WindContrGalCR_2024}
Peron, G., Casanova, S., Gabici, S., Baghmanyan, V., \& Aharonian, F. 2024,
  Nature Astronomy

\bibitem[{{Piskunov} {et~al.}(2018){Piskunov}, {Just}, {Kharchenko}, {Berczik},
  {Scholz}, {Reffert}, \& {Yen}}]{Piskunov_GalSCGlobSurvIV_2018}
{Piskunov}, A.~E., {Just}, A., {Kharchenko}, N.~V., {et~al.} 2018, \aap, 614,
  A22

\bibitem[{{Planck Collaboration} {et~al.}(2011){Planck Collaboration}, {Ade},
  {Aghanim}, {Arnaud}, {Ashdown}, {Aumont}, {Baccigalupi}, {Balbi}, {Banday},
  {Barreiro}, {Bartlett}, {Battaner}, {Benabed}, {Beno{\^\i}t}, {Bernard},
  {Bersanelli}, {Bhatia}, {Bock}, {Bonaldi}, {Bond}, {Borrill}, {Bouchet},
  {Boulanger}, {Bucher}, {Burigana}, {Cabella}, {Cardoso}, {Catalano},
  {Cay{\'o}n}, {Challinor}, {Chamballu}, {Chiang}, {Chiang}, {Christensen},
  {Clements}, {Colombi}, {Couchot}, {Coulais}, {Crill}, {Cuttaia}, {Dame},
  {Danese}, {Davies}, {Davis}, {de Bernardis}, {de Gasperis}, {de Rosa}, {de
  Zotti}, {Delabrouille}, {Delouis}, {D{\'e}sert}, {Dickinson}, {Dobashi},
  {Donzelli}, {Dor{\'e}}, {D{\"o}rl}, {Douspis}, {Dupac}, {Efstathiou},
  {En{\ss}lin}, {Eriksen}, {Falgarone}, {Finelli}, {Forni}, {Fosalba},
  {Frailis}, {Franceschi}, {Fukui}, {Galeotta}, {Ganga}, {Giard}, {Giardino},
  {Giraud-H{\'e}raud}, {Gonz{\'a}lez-Nuevo}, {G{\'o}rski}, {Gratton},
  {Gregorio}, {Grenier}, {Gruppuso}, {Hansen}, {Harrison}, {Helou},
  {Henrot-Versill{\'e}}, {Herranz}, {Hildebrandt}, {Hivon}, {Hobson}, {Holmes},
  {Hovest}, {Hoyland}, {Huffenberger}, {Jaffe}, {Jones}, {Juvela}, {Kawamura},
  {Keih{\"a}nen}, {Keskitalo}, {Kisner}, {Kneissl}, {Knox}, {Kurki-Suonio},
  {Lagache}, {Lamarre}, {Lasenby}, {Laureijs}, {Lawrence}, {Leach}, {Leonardi},
  {Leroy}, {Lilje}, {Linden-V{\o}rnle}, {L{\'o}pez-Caniego}, {Lubin},
  {Mac{\'\i}as-P{\'e}rez}, {MacTavish}, {Maffei}, {Maino}, {Mandolesi}, {Mann},
  {Maris}, {Martin}, {Mart{\'\i}nez-Gonz{\'a}lez}, {Masi}, {Matarrese},
  {Matthai}, {Mazzotta}, {McGehee}, {Meinhold}, {Melchiorri}, {Mendes},
  {Mennella}, {Miville-Desch{\^e}nes}, {Moneti}, {Montier}, {Morgante},
  {Mortlock}, {Munshi}, {Murphy}, {Naselsky}, {Natoli}, {Netterfield},
  {N{\o}rgaard-Nielsen}, {Noviello}, {Novikov}, {Novikov}, {O'Dwyer}, {Onishi},
  {Osborne}, {Pajot}, {Paladini}, {Paradis}, {Pasian}, {Patanchon},
  {Perdereau}, {Perotto}, {Perrotta}, {Piacentini}, {Piat}, {Plaszczynski},
  {Pointecouteau}, {Polenta}, {Ponthieu}, {Poutanen}, {Pr{\'e}zeau}, {Prunet},
  {Puget}, {Reach}, {Reinecke}, {Renault}, {Ricciardi}, {Riller},
  {Ristorcelli}, {Rocha}, {Rosset}, {Rowan-Robinson}, {Rubi{\~n}o-Mart{\'\i}n},
  {Rusholme}, {Sandri}, {Santos}, {Savini}, {Scott}, {Seiffert}, {Shellard},
  {Smoot}, {Starck}, {Stivoli}, {Stolyarov}, {Stompor}, {Sudiwala}, {Sygnet},
  {Tauber}, {Terenzi}, {Toffolatti}, {Tomasi}, {Torre}, {Tristram}, {Tuovinen},
  {Umana}, {Valenziano}, {Vielva}, {Villa}, {Vittorio}, {Wade}, {Wandelt},
  {Wilkinson}, {Yvon}, {Zacchei}, \& {Zonca}}]{Planck2011}
{Planck Collaboration}, {Ade}, P.~A.~R., {Aghanim}, N., {et~al.} 2011, \aap,
  536, A19

\bibitem[{{Reid} {et~al.}(2019){Reid}, {Menten}, {Brunthaler}, {Zheng}, {Dame},
  {Xu}, {Li}, {Sakai}, {Wu}, {Immer}, {Zhang}, {Sanna}, {Moscadelli}, {Rygl},
  {Bartkiewicz}, {Hu}, {Quiroga-Nu{\~n}ez}, \& {van
  Langevelde}}]{Reid_GalRotCurve_2019}
{Reid}, M.~J., {Menten}, K.~M., {Brunthaler}, A., {et~al.} 2019, \apj, 885, 131

\bibitem[{Reimer {et~al.}(2006)Reimer, Pohl, \&
  Reimer}]{Reimer_CRfromWWinteraction_2006}
Reimer, A., Pohl, M., \& Reimer, O. 2006, The Astrophysical Journal, 644, 1118

\bibitem[{{Rosslowe} \& {Crowther}(2015)}]{Rosslowe_WRsAge_2015}
{Rosslowe}, C.~K. \& {Crowther}, P.~A. 2015, \mnras, 447, 2322

\bibitem[{{Schaerer} \& {Maeder}(1992)}]{Schaerer_WRsMLR_1992}
{Schaerer}, D. \& {Maeder}, A. 1992, \aap, 263, 129

\bibitem[{{Silich} \& {Tenorio-Tagle}(2013)}]{Silich_RadPressYMSCs_2013}
{Silich}, S. \& {Tenorio-Tagle}, G. 2013, \apj, 765, 43

\bibitem[{{Strong} {et~al.}(2009){Strong}, {Moskalenko}, {Porter},
  {J{\'o}hannesson}, {Orlando}, \& {Digel}}]{Strong_Galprop_2009}
{Strong}, A.~W., {Moskalenko}, I.~V., {Porter}, T.~A., {et~al.} 2009, arXiv
  e-prints, arXiv:0907.0559

\bibitem[{{Strong} {et~al.}(2000){Strong}, {Moskalenko}, \&
  {Reimer}}]{Strong_GalpropDiffGamma_2000}
{Strong}, A.~W., {Moskalenko}, I.~V., \& {Reimer}, O. 2000, \apj, 537, 763

\bibitem[{{Vieu} {et~al.}(2022){Vieu}, {Gabici}, {Tatischeff}, \&
  {Ravikularaman}}]{Vieu_MSC+SNR_2022}
{Vieu}, T., {Gabici}, S., {Tatischeff}, V., \& {Ravikularaman}, S. 2022,
  \mnras, 512, 1275

\bibitem[{{Wakely} \& {Horan}(2008)}]{Wakely_TeVCat_2008}
{Wakely}, S.~P. \& {Horan}, D. 2008, in International Cosmic Ray Conference,
  Vol.~3, International Cosmic Ray Conference, 1341--1344

\bibitem[{{Weaver} {et~al.}(1977){Weaver}, {McCray}, {Castor}, {Shapiro}, \&
  {Moore}}]{Weaver_1977}
{Weaver}, R., {McCray}, R., {Castor}, J., {Shapiro}, P., \& {Moore}, R. 1977,
  \apj, 218, 377

\bibitem[{{Wenger} {et~al.}(2018){Wenger}, {Balser}, {Anderson}, \&
  {Bania}}]{Wenger_KinDistCode_2018}
{Wenger}, T.~V., {Balser}, D.~S., {Anderson}, L.~D., \& {Bania}, T.~M. 2018,
  \apj, 856, 52

\bibitem[{{Yang} \& {Aharonian}(2017)}]{Yang2017NGC3603}
{Yang}, R.-z. \& {Aharonian}, F. 2017, \aap, 600, A107

\bibitem[{{Yang} {et~al.}(2018){Yang}, {de O{\~n}a Wilhelmi}, \&
  {Aharonian}}]{Yang_Wd2Fermi_2018}
{Yang}, R.-z., {de O{\~n}a Wilhelmi}, E., \& {Aharonian}, F. 2018, \aap, 611,
  A77

\bibitem[{{Zhang} {et~al.}(2023){Zhang}, {Huang}, {Xu}, {Zhao}, \&
  {Yuan}}]{Zhang_LHAASOandFermiLATDiffuseGamma_2023}
{Zhang}, R., {Huang}, X., {Xu}, Z.-H., {Zhao}, S., \& {Yuan}, Q. 2023, arXiv
  e-prints, arXiv:2305.06948

\bibitem[{{Zinnecker} \& {Yorke}(2007)}]{Zinnecker_MstarMax150_2007}
{Zinnecker}, H. \& {Yorke}, H.~W. 2007, \araa, 45, 481

\end{thebibliography}

\appendix
\section{Generation of the mock stellar population} \label{app:f_nstar}

When generating the mock stellar population in a cluster, it is operationally more straightforward to use the number of stars rather than the cluster mass. These two quantities are related by:
\begin{equation}
    N_\star=\Lambda M_{\rm sc}\, ,
    \label{eq:LambdaMsc-Nstar1}
\end{equation}
with the parameter $\Lambda$ defined as:
\begin{equation}
\label{eq:LambdaMsc-Nstar}
    \Lambda=\frac{\int_{M_{\star,\min}}^{M_{\star,\max}} f_\star(M_\star)dM_\star}{\int_{M_{\star,\min}}^{M_{\star,\max}} M_\star f_\star (M_\star) dM_\star} \, ,
\end{equation}
where $M_\star$ is the stellar mass, $f_\star$ is the initial stellar mass function, and $M_{\star,\min}$ and $M_{\star,\max}$ are, respectively, the minimum and maximum stellar masses that can be generated in a cluster.

Because the number of stars in a cluster is the starting ingredient to generate the mock stellar population, for convenience, we operationally build up the cluster population in the Galaxy not as a function of the cluster mass, but rather as a function of $N_\star$. This implies that the distribution function used to extract the clusters is $f'(N_\star) \equiv \frac{dN_{\rm sc}}{dN_{\star}}$ and is related to $f(M_{\rm sc})$ as a rewriting of the initial cluster mass function in terms of $N_\star$ using Equation~\ref{eq:LambdaMsc-Nstar1}:
\begin{equation}
    f'(N_\star) = f(M_{\rm sc}) \frac{d M_{\rm sc}}{d N_{\star}} = 1.2 \left( \Lambda M_{\odot} \right)^{0.54} N_\star^{-1.54} \,.
\end{equation}
By doing so, the maximum and minimum cluster mass directly translate to a maximum and minimum number of stars, i.e. $N_{\star,\min}=\Lambda M_{\rm sc, \min}$ and $N_{\star,\max}=\Lambda M_{\rm sc, \max}$.

\section{Definition of masks}
\label{app:mask}
The measurements of the Galactic diffuse emission in both \texttt{ROI}s are provided after masking known very-high-energy $\gamma$-ray sources. Therefore, in order to compare the emission from simulated YMSCs with the available data, our simulated sources should be masked following a similar criterion. 
The masks used by \cite{Zhang_LHAASOandFermiLATDiffuseGamma_2023} and \cite{Cao_LHAASODiffuseGamma_2023} (M$1_{\rm L}$ for \texttt{ROI1} and M$2_{\rm L}$ for \texttt{ROI2} in the following) are built by
removing the sources reported in the TeVcat plus the ones recently discovered by the KM2A detector of LHAASO \citep{Cao_LHAASOCat_2023}. 
These masks cannot be applied to our simulated maps by simply using their spatial coordinates, because our sources are in different positions from the real ones and different from one realization to the other. The method we implemented to mimic the masks used by \cite{Cao_LHAASODiffuseGamma_2023} is described in the following. 

We used three different kinds of masks. The first mask, $\mathcal{A}$, is designed to remove regions of the Galactic Plane where, on average, most of the galactic sources are located. Following the shape of M$1_{\rm L}$, one can easily see that, for $l\lesssim 70^\circ$, a large fraction of the Galactic Plane is masked. We hence define $\mathcal{A}$ so as to remove the zone of the sky defined between $15^\circ<l<70^\circ$ and $|b|<1.5^\circ$. We also remove the region corresponding to the projected size of the Local Spiral arm, as an over-density of sources is expected there. Such a region corresponds to a circular area centered on $(l=73.5^\circ, \, b=0^\circ)$ with radius $6.5^\circ$. Notice that in \texttt{ROI2} a sparse distribution of the galactic sources is expected, as the outer region of the Milky Way is less crowded. Consequently, there is not a specific zone of the plane to mask. Hence, we define this specific type of mask only for \texttt{ROI1}.

\begin{figure*}[t]
\begin{center}
\includegraphics[width=0.95\textwidth]{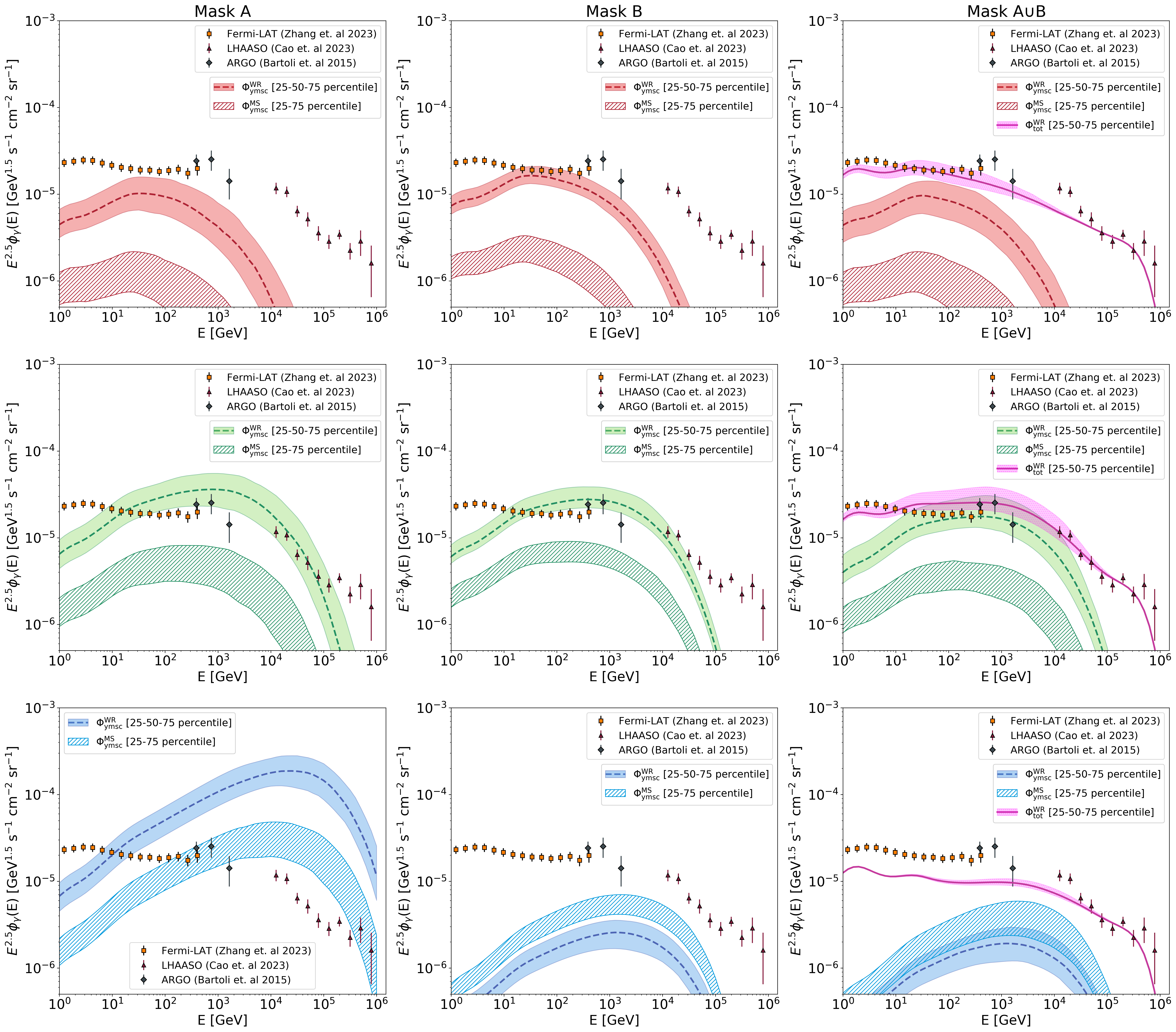} 
\caption{Contribution to the diffuse $\gamma$-ray emission from a synthetic population of YMSCs in \texttt{ROI1} for the three different cases of diffusion coefficient considered: Kolmogorov (upper row), Kraichnan (central row) and Bohm-like (lower row). Each column shows the flux for the three masks implemented: $\mathcal{A}$ (left column), $\mathcal{B}$ (central column) and $\mathcal{A}\cup \mathcal{B}$ (right column). In each plot, the dashed line represents the median flux per energy bin after considering 100 different realisations of the Milky Way population of YMSCs, while the associated shaded region encloses the 25-75 percentile flux. The striped region instead encompasses the 25-75 percentile flux when WR stars are not included in the stellar mock populations. The violet solid lines in the right column and their associated shaded regions show the median and the 25-75 percentile diffuse $\gamma$-ray flux after the addition of emission from the CR sea.}
\label{fig:ROI1}
\end{center}
\end{figure*}

The second type of masks, $\mathcal{B}$, is defined so as to remove the emission from those YMSCs that would be detectable by the LHAASO experiment. 
To do so, we mask the positions of the sky where the significance of the surface brightness at 100 TeV ($\mathcal{S}_{100}$) exceeds by 5 $\sigma$ the brightness of the measured galactic diffuse emission, considered here as the nominal background. We use the emission at 100~TeV as it correspond to the energy where LHAASO has the best sensitivity \citep{Cao_LHAASOScienceBook_2019}. Hereafter, the subscript "100" indicates that the quantity is evaluated at 100~TeV.
At each position of the sky belonging to the \texttt{ROI}s with coordinates ($l_0$, $b_0$), the significance of the emission is calculated as:
\begin{equation}
\label{eq:significance}
    \mathcal{S}_{100}(l_0, b_0)=\iint \Theta(l_0, b_0, r_{100}^{\rm psf})\frac{N_{100}^{\rm sc}(l, b)}{\sqrt{N_{100}^{\rm gde}(l, b)}} dl\, db.
\end{equation}
A given position of the sky is then masked if $\mathcal{S}_{100}>5$. In Equation~\ref{eq:significance}, $\Theta(l_0, b_0, r_{100}^{\rm psf})$ is a 2D top hat function centered on ($l_0$, $b_0$) with radius equal to the LHAASO point spread function ($r_{100}^{\rm psf}=0.2^\circ$) \citep{Cao_LHAASOScienceBook_2019}, while $N_{100}^{\rm sc}(l, b)$ and $N_{100}^{\rm gde}(l, b)$ are the expected counts from the YMSCs and the galactic diffuse $\gamma$-ray emission, respectively. The latter quantities are calculated as:
\begin{subequations}
    \begin{equation}
        N_{100}^{\rm sc}(l,b)=\phi_{100}^{\rm sc}(l,b)\, \epsilon_{100}(l,b)\, \Delta E_{100}
    \end{equation}
    \begin{equation}
        N_{100}^{\rm gde}(l,b)=\phi_{100}^{\rm bkg}\, \epsilon_{100}(l,b)\, \Delta E_{100} \, ,
    \end{equation}  
\end{subequations}
where $\phi_{100}^{\rm sc}(l,b)$ is the flux at the position ($l$, $b$) only due to YMSCs, while $\phi_{100}^{\rm gde}$ is the measured value of the diffuse galactic flux measured by LHAASO \citep{Cao_LHAASODiffuseGamma_2023}. Note that by doing so, we are assuming that the observed Galactic diffuse $\gamma$-ray emission is constant over the entire \texttt{ROI}. The value $\Delta E_{100}$ is the energy bin width corresponding to the measured flux point at 100~TeV \citep{Cao_LHAASODiffuseGamma_2023}. Finally, $\epsilon_{100}(l,b)$ is the detector exposure at the position ($l$, $b$), which is calculated as:
\begin{equation}
    \epsilon_{100}(l,b)=A_{\rm eff, 100}\, t_{\rm exp}(l, b)
\end{equation}
where $A_{\rm eff, 100}$ is the KM2A effective area for $\gamma$-like events \citep{Cao_LHAASOScienceBook_2019} and $t_{\rm exp}(l, b)$ is the exposure time of a given location in the sky. For the latter we consider that the detector array has operated also during the construction phase\footnote{Because a significant portion of the observations were obtained with the array under construction, the detector sensitivity is not constant in time. We account for this by assuming that the live-time scales linearly with the active percentage of the detector.}, so that
\begin{equation}
    t_{\rm exp}(l, b)=\left [\left (\frac{1}{2}302+\frac{3}{4}219+423 \right )\, \rm{days} \right]\times t_{\rm hpd}(l,b)
\end{equation}
where $t_{\rm hpd}(l,b)$ is the amount of time per day a given sky position is found at zenith angles ($\zeta$) below $50^\circ$, which can be evaluated as:
\begin{equation}
    t_{\rm hpd}(l,b)=\int_0^{24\, \rm h} \mathcal{H}\left (50^\circ-\zeta(l,b,t)\right)\cos[\zeta(l,b,t)] dt \, ,
\end{equation}
where $\mathcal{H}\left (50^\circ-\zeta(l,b,t)\right)$ is the Heaviside function, and
\begin{eqnarray}
\label{eq:cosZeta}
  \cos[\zeta(l,b,t)] = \sin(\delta_{\rm L}) \, \sin[\delta(l,b)] \hspace{2.5cm}\nonumber \\ 
  + \cos(\delta_{\rm L}) \, \cos[\delta(l,b)] \, \cos\left[ 15^\circ\left(\frac{t}{\rm h}\right)\right] \, .
\end{eqnarray}
In Equation~\eqref{eq:cosZeta} the angles $\delta_{\rm L}$ and $\delta(l, b)$ are the equatorial declination of the LHAASO site and the given position of the sky, respectively. Finally, the last type of mask, $\mathcal{A} \cup \mathcal{B}$, is obtained by merging $\mathcal{A}$ with $\mathcal{B}$. In \texttt{ROI1} the $\mathcal{A} \cup \mathcal{B}$ mask is the one that best reproduces M$1_{\rm L}$. Figure~\ref{fig:ROI1} shows how the contribution to the diffuse gamma-ray flux in ROI1 changes after applying the $\mathcal{A}$ (left column), $\mathcal{B}$ (central column) and $\mathcal{A} \cup \mathcal{B}$ (right column) masks.

\begin{figure*}[t]
\begin{center}
\includegraphics[width=0.95\textwidth]{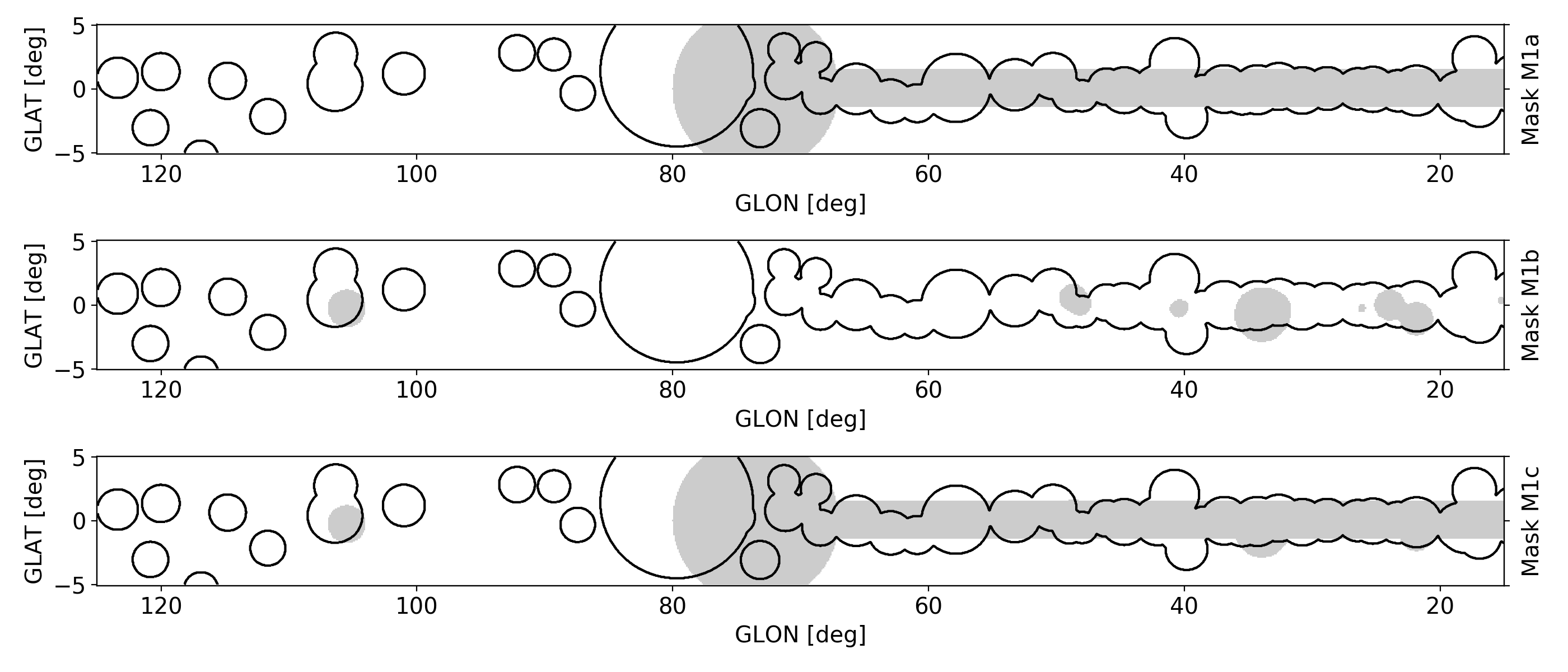} 
\caption{An example of the three masks employed to estimate the diffuse $\gamma$-ray emission in \texttt{ROI1}. Upper panel: mask $\mathcal{A}$ used to exclude the Local Spiral Arm and a large fraction of the inner Galactic Plane where most of the $\gamma$-ray sources are expected to be located. Central panel: mask $\mathcal{B}$ used to remove the emission from the most luminous clusters that should have been detected by LHAASO (this example mask is the one built based for the specific source realization shown in Figure~\ref{fig:SC_distrib} and assuming the Kraichnan diffusion coefficient). Lower panel: mask $\mathcal{A} \cup \mathcal{B}$. For comparison, in all panels, the M$1_{\rm L}$ mask is shown as black contours.}
\label{fig:masks_comparison}
\end{center}
\end{figure*}

Figure~\ref{fig:masks_comparison} shows an example of how $\mathcal{A}$, $\mathcal{B}$, and $\mathcal{A} \cup \mathcal{B}$ compare with M$1_{\rm L}$. This specific configuration refers to the Galaxy realization showed in Figure~\ref{fig:SC_distrib} with the assumption of Kraichnan diffusion and inclusion of the WR contribution.
Using all the 100 different realisations of the Milky Way, we can quantitatively estimate the difference between our masks and the original ones by evaluating the median ratio of the non-masked areas. For the $\mathcal{A} \cup \mathcal{B}$ mask we found this ratio to be 1.22, 1.19 and 1.1 for the Kolmogorov, Kraichnan and Bohm cases, respectively. 
The same ratios calculated for the \texttt{ROI2} are 1.24, 1.24 and 1.22. Those numbers are estimated including WR stars in the stellar population. We found no significant difference when WRs are excluded. In summary, our sky regions is between 10\% and 20\% larger than the one considered by \cite{Cao_LHAASODiffuseGamma_2023}.

\begin{figure*}[t]
\begin{center}
\includegraphics[width=0.95\textwidth]{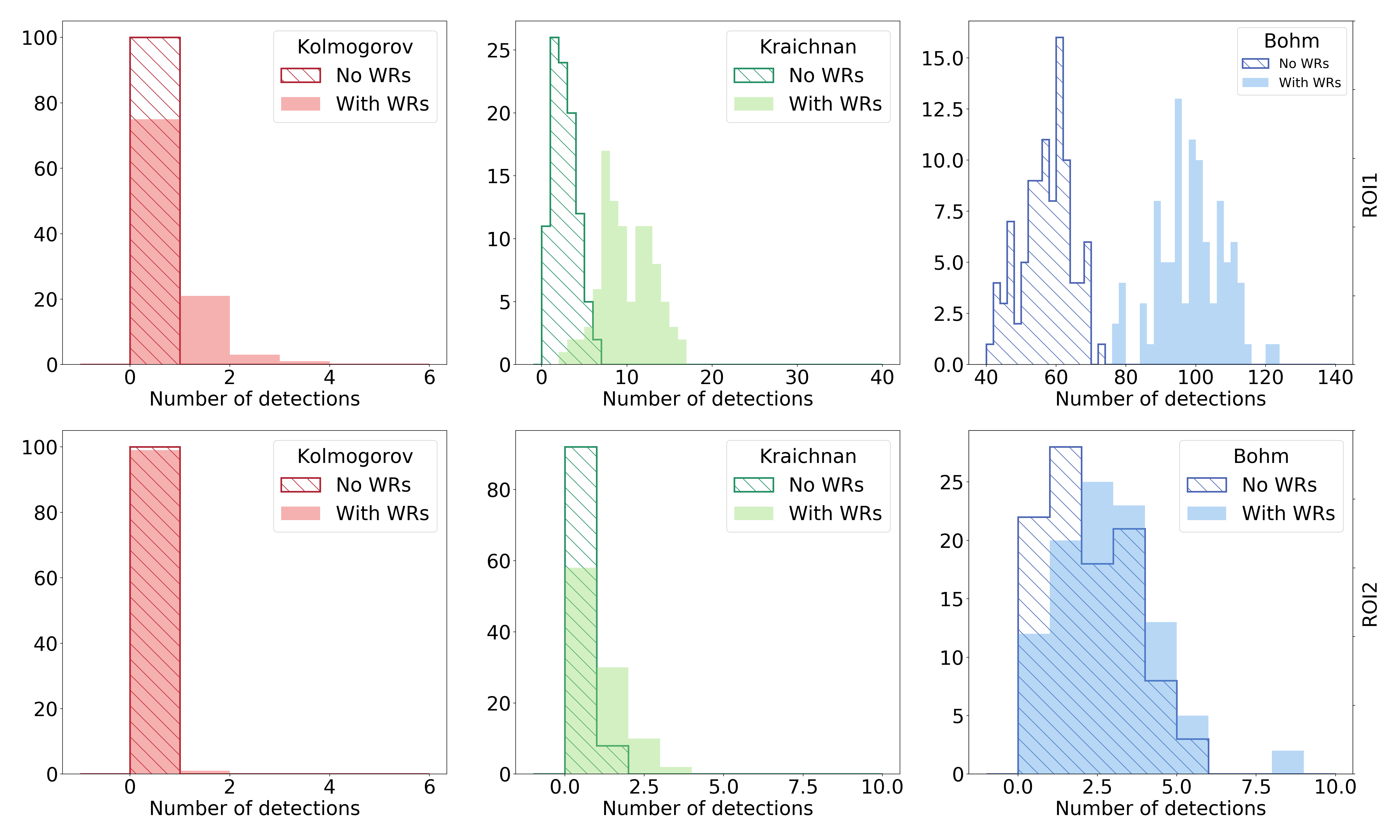}
\caption{Distribution of the number of synthetic YMSCs detectable by LHAASO in \texttt{ROI1} (upper row) and \texttt{ROI2} (bottom row) for the three diffusion coefficients considered in this analysis: Kolmogorov, Kraichnan and Bohm (from left to right). We label a YMSC as detected if the significance of its emission at 100~TeV is larger than 5$\sigma$. Color filled (striped)  histograms correspond to the scenarios where WRs are (are not) included in the mock stellar populations.}
\label{fig:Ndetect}
\end{center}
\end{figure*}

The same method employed to generate the mask $\mathcal{B}$ can also be used to estimate the number of YMSCs that should have been detected by LHAASO. This can be seen in Figure~\ref{fig:Ndetect} were the distribution of the number of detected clusters over the 100 different realisations is shown. Note that this number of detections must be considered as an order of magnitude estimate, as the detectability of a cluster should be evaluated after considering the instrument sensitivity over the entire energy range. Interestingly, the number of detections for the Kolmogorov and Kraichnan cases is consistent with the number of sources in the LHAASO catalog that can be associated to SCs \citep{Mitchell+2024}. In contrast, the predicted median number of YMSCs in the Bohm case ($\sim$ 100 if WRs are included in the stellar populations) is of the same order of the total number of sources in the LHAASO catalog, suggesting that this scenario is likely to be too extreme. 

\section{Diffuse emission from the CR sea}
\label{app:CRseaFlux}
The $\gamma$-ray spectrum produced by the CR sea is calculated as in the work by \cite{Peron2022}, namely by assuming that the CR spectrum around the Galaxy is the same as the one locally measured at Earth. We consider the spectrum of CR protons, $J(E_p)$, measured by AMS-02, Dampe, Kascade, and IceTop and fitted by \cite{Lipari2019}. To account for heavier elements, both in the CR composition and in the target, we consider a nuclear enhancement factor, $\xi_N$, computed following \cite{Mori2009}, by considering  a solar-like composition for the target and the different measured energy spectra for the different CR elements. This implies that the nuclear enhancement factor depends on energy. 
We used the differential cross section given by \cite{Kafexhiu_SigmaPi0Gamma_2014}, choosing the SYBILL parametrization. Potential differences arising from the different choice of the cross-section parametrization are neglected in this work. The CR flux can be then computed as:
\begin{equation}
   F_{\gamma}(E_\gamma)= N_H \, \Delta \Omega \, \xi_N(E_{\gamma}) \int d E_p \frac{d\sigma(E_{\gamma},E_p)}{dE_p} J(E_p) .
\end{equation}
 
The density of the target, $N_H$, is derived from the Planck dust-opacity, $\tau_D$, observed in the regions outside the M$1_{\rm L}$ and M$2_{\rm L}$ masks. The result is $N_H =  7.85 \times 10^{21}$ cm$^{-2}$ in \texttt{ROI1} and $N_H=5.89 \times 10^{21}$ cm$^{-2}$ in \texttt{ROI2}, when assuming a dust-to-H conversion factor $X_{D}^{-1} \equiv \frac{\tau_D}{N_H} = 1.18\times 10^{-26} \mathrm{cm}^{2}$ \citep{Planck2011}. Finally, we consider the fact that the $\gamma$-ray emitting gas is located at different distances from us, and consequently, its emission can be affected by absorption due to the $\gamma$-$\gamma$ interaction between the high-energy photons and the interstellar radiation fields. The effect of absorption is evaluated for the average Galactic radiation fields provided by \cite{Lipari2018} as a function of the source distance. We here derive the gas distance by considering the measured radial velocity at each position. To do this we used the gas maps provided along with the \textit{Galprop} code \citep{Strong_Galprop_2009}, already decomposed according to distance from the Galactic Center. This allows us to estimate what fraction of column density is located at which distance and weight for the interstellar absorption accordingly. 

\end{document}